\documentclass[twocolumn,aps,superscriptaddress,longbibliography]{revtex4-1}
\usepackage{graphicx}
\usepackage{dcolumn}
\usepackage{mathtools}
\usepackage{romannum}
\usepackage{simplewick}
\usepackage{bm}
\usepackage{rotating}
\usepackage{adjustbox}
\usepackage{hyperref}
\usepackage[usenames,dvipsnames]{color}
\usepackage{amsmath}
\usepackage{amssymb}
\usepackage{times}

\DeclareUnicodeCharacter{2212}{-}
\begin{document}

\title{FSRCC two-valence calculations of clock transition properties, 
dipole polarizability and isotope shifts in Fermionic and Bosonic Sr}

\author{Palki Gakkhar}
\affiliation{Department of Physics, Indian Institute of Technology,
             Hauz Khas, New Delhi 110016, India}


\author{D. Angom}
\affiliation{Department of Physics, Manipur University,
             Canchipur 795003, Manipur, India}

\author{B. K. Mani}
\email{bkmani@physics.iitd.ac.in}
\affiliation{Department of Physics, Indian Institute of Technology,
             Hauz Khas, New Delhi 110016, India}             

\begin{abstract}

We employ an all-particle multireference Fock-space relativistic coupled-cluster
(FSRCC) theory to study the $5s^2{\;^1}S_0 - 5s5p{\;^3P^o_0}$ clock transition
in both Fermionic and Bosonic isotopes of Sr. We compute the excitation energies
for several low-lying states, E1 and M1 transition amplitudes, hyperfine reduced
matrix elements, and isotope shifts using FSRCC theory.
Further, using our results on E1, M1 and HFS reduced matrix elements,
we calculate the lifetime of the metastable
clock states for $^{87}$Sr and $^{88}$Sr. Furthermore, we employ perturbed
relativistic coupled-cluster (PRCC) theory to compute the ground state
electric dipole polarizability of Sr. To improve the accuracy of our results,
we incorporate the corrections from the relativistic and quantum
electrodynamical (QED) effects, and perturbative triples to all our
calculations. Moreover, we employ large bases to ensure the convergence
of the computed properties.
Our computed excitation energies are in good agreement with the experimental
data for low-lying excited states. Our results for E1, M1 and HFS reduced
matrix elements are within the experimental error bars, however, with slight
difference from the previous calculations due to more accurate treatment of
electron correlations in our calculation. Our computed lifetime of the clock
state for $^{87}$Sr is within the error bars of the available experimental
results, whereas for $^{88}$Sr it is an order of magnitude smaller than the 
only available calculation [R. Santra {\em et al.,} Phys. Rev. A {\bf 69}, 042510 (2004)]
using model potential. Our PRCC result for the ground state polarizability is 
in good agreement with the experiment, and smaller than previous calculations.
As can be expected, our FSRCC results on isotope shift parameters show 
differences from the MCDF calculations.
From the detailed analysis of the results, we find that the corrections
from the Breit interaction, QED effects, and perturbative triples are
crucial to get accurate clock transition properties in Sr.
Moreover, {\em valence-valence} electron correlation is important to
get accurate energies and properties of Sr.
\end{abstract}



\maketitle

\section{Introduction}

The unprecedented accuracy of optical atomic clocks has revolutionized 
the precision timekeeping which has led to several fundamental as well as 
technological implications across different fields \cite{ludlow-15, subhadeep-23}.
By using ultra-narrow optical transitions in atoms or ions, these clocks achieve
high accuracy and stability, surpassing the traditional
microwave frequency based clocks. This high level of precision enables new
possibilities for testing the fundamental physics beyond the standard 
model (BSM) \cite{dzuba-18, berengut-18}, including the variation
in the fundamental constants \cite{safronova-19, prestage-95, karshenboim-10}
and the tests of general relativity \cite{kolkowitz-16, bondarescu-12}.
Beyond fundamental physics, optical atomic clocks are also useful in practical
applications such as geodesy \cite{mehlstaubler_18}, quantum computers \cite{weiss-17, wineland-02},
redefining the SI unit of second \cite{riehle-15} and navigation systems \cite{grewal-13, major-13}.
Their role in advancing technologies for radar, radio astronomy, and 
telecommunications \cite{grewal-13} further underscores their 
importance in technological applications.

Among the leading candidates for optical atomic clocks, neutral atoms and singly
charged ions based clocks are at the forefront of precision timekeeping.
The ion-based clocks
such as $^{27}\text{Al}^+$ \cite{marshall-25}, $^{40}\text{Ca}^+$ \cite{chwalla-09},
$^{88}\text{Sr}^+$ \cite{lindwall-25}, $^{115}\text{In}^+$ \cite{ohtsubo-20}, $^{171}\text{Yb}^+$ \cite{dorscher-21}, 
and $^{199}\text{Hg}^+$ \cite{diddams-01}
are widely studied due to their high precision \cite{subhadeep-23}.
Among these, $^{27}{\rm Al}^+$ quantum logic clock currently holds
the record for the highest accuracy, with a fractional frequency uncertainty
of \(5.5 \times 10^{-19}\) \cite{marshall-25}. Its hyperfine induced 
$3s^2\ ^1S_0 (F = 5/2) \rightarrow 3s3p\ ^3P_0 (F = 5/2)$ clock transition 
is reported to demonstrate a significant advancement in precision timekeeping.
Among the neutral atoms clocks, lattice clocks based on $^{87,88}\text{Sr}$ \cite{aeppli-24, baillard-07}, 
$^{171}\text{Yb}$ \cite{mcgrew-18} and $^{199}\text{Hg}$ \cite{yamanaka-15} have gained prominence 
in recent years \cite{subhadeep-23}.
These clocks have demonstrated a remarkable accuracy and stability, pushing
the boundaries of precision timekeeping.
For instance, the hyperfine induced 
$5s^2\ ^1S_0 (F = 9/2) \rightarrow 5s5p\ ^3P_0 (F = 9/2)$ transition 
in Fermionic Sr ($^{87}{\rm Sr}$) is reported to show an excellent balance 
between accuracy and practicality with a fractional frequency 
shift of \(8.1 \times 10^{-19}\) \cite{aeppli-24}.
It is, perhaps, the most accurate lattice clock in existence today.
It leverages thousands of neutral atoms trapped in an optical lattice, providing
a superior short-term stability and low fractional frequency shifts due to
environmental perturbations.
Moreover, compared to Al\(^+\) clocks, which face practical challenges 
due to complex experimental setups and limited short-term 
stability\cite{schmidt-05, chou-10}, \(^{87}\text{Sr}\)
clocks require relatively simpler experimental setups \cite{poli-09}.
In addition, the availability of series of stable isotopes allows to probe the
nuclear properties and the search for the new physics beyond the 
standard model \cite{delaunay-17}. While studying isotope shifts
between a Fermionic and Bosonic isotopes offers a unique insights into 
nuclear properties, examining isotope shifts among Bosonic 
isotopes ($^{84}{\rm Sr}$, $^{86}{\rm Sr}$, $^{88}{\rm Sr}$) could provide
important insights into the BSM physics \cite{delaunay-17}.
The {\em two} photon E1+M1 allowed $5s^2\ ^1S_0 \rightarrow 5s5p\ ^3P_0$
clock transition in Bosonic Sr ($^{88}{\rm Sr}$) is demonstrated to be an
excellent candidate for optical clock, particularly for space-based
applications \cite{origlia-18}. The advantages with $^{88}{\rm Sr}$
over $^{87}{\rm Sr}$ are, a higher natural abundance (83\%, as compared
to only 7\% for \(^{87}\text{Sr}\)) and the absence of hyperfine
splitting, which simplifies the cooling and spectroscopic 
processes \cite{origlia-18}. These make $^{88}{\rm Sr}$ a suitable 
candidate for a robust and transportable optical lattice clock.
Notably, experiments conducted with $^{88}\text{Sr}$ have shown
remarkable performance, achieving an instability
of \(3 \times 10^{-18}\) and a fractional frequency shift
of \(2 \times 10^{-17}\) \cite{origlia-18}.

Despite this important prospect with Sr as an optical clock, its clock
transition properties have not been computed accurately. For example,
for \(^{88}\text{Sr}\), to the best of our knowledge, there is only one
study on the lifetime of the metastable clock state and
that too is obtained using a model potential based calculation \cite{santra-04}.
Considering that there are no experimental data, fully {\em ab initio} based
calculations, especially using an accurate many-body methods such as 
relativistic coupled-cluster (RCC), would be crucial to get better insights
into the clock transition. Similarly, for $^{87}{\rm Sr}$, the available 
theoretical data on the lifetime of the clock state are limited
to model potential \cite{santra-04}, MCDF \cite{lu-23} and CI+MBPT \cite{porsev-04}
calculations. Moreover, the inclusion of relativistic and QED corrections 
in the properties calculation is essential to obtain reliable results.
It can thus be surmised that there is a clear gap in terms of the 
availability of accurate properties results for clock 
transition in Sr.

Considering this, in this work, we employ an all-particle multireference
Fock-space relativistic coupled-cluster (FSRCC) theory to investigate the
clock transition properties in both Fermionic and Bosonic isotopes of Sr.
It should be noted that, RCC theory is one of the most reliable many-body
methods for atomic structure calculations, as it accounts for electron correlation
to all orders of residual Coulomb interaction. The RCC has been employed
to obtain accurate properties results in several closed-shell and one-valence atoms
and ions \cite{pal-07,mani-09,nataraj-11,ravi-20}. Its application for two-valence
atomic systems, is, however, limited to few studies \cite{eliav-95, eliav-95b, mani-11b, ravi-21, palki-24}
due to the complications associated with the implementation of FSRCC theory for
multi-reference systems \cite{eliav-95,eliav-95b,mani-11b,ravi-21}.
Using the FSRCC theory, we have computed the excitation energies for several
low lying states, E1 and M1 transition amplitudes, and hyperfine matrix elements
associated with $5s^2\ ^1S_0 \rightarrow 5s5p\ ^3P_0$ clock transition.
Further, using these results, we have calculated the lifetimes of the metastable
clock states for both $^{87}{\rm Sr}$ and $^{88}{\rm Sr}$.
In addition, we employ FSRCC theory to calculate the normal mass shift (NMS),
specific mass shift (SMS), and field shift (FS) parameters for clock transition 
$5s^2\ ^1S_0 \rightarrow 5s5p\ ^3P_0$ and two intercombination 
lines, $5s^2\ ^1S_0 \rightarrow 5s5p\ ^3P_1$ 
and $5s^2\ ^1S_0 \rightarrow 5s5p\ ^1P_1$, in Sr. 
Furthermore, we employ a perturbed relativistic coupled-cluster (PRCC)
theory \cite{chattopadhyay-12, chattopadhyay-14, 
chattopadhyay-15, ravi-20, ravi-21b} to calculate the ground state electric
dipole polarizability ($\alpha$) of Sr. It should be noted that, $\alpha$
is a crucial parameter in estimating the blackbody radiation (BBR) shift in
the clock transition frequency. The other key merits of our work are: first,
to further improve the accuracies of our results, we have incorporated the
corrections from the Breit interaction, QED effects, and perturbative triples
in our calculations; second, to test the convergence of our results, we have
employed large basis sets in the FSRCC and PRCC calculations.

The remainder of the paper is organized into five sections. In Sec. II, we
provide a brief description of the FSRCC theory for two-valence atomic systems.
We have given the coupled-cluster working equation for two-valence systems.
In the same section, we also provide the expressions for hyperfine induced
and E1M1 transition rates, and perturbative triples corrections to the
energy of two-valence systems. The results obtained from our calculations are
presented and discussed in different subsections of Sec. III.
Theoretical uncertainty in our computed results is discussed
in Sec. IV of the paper. Unless stated otherwise, all results and
equations presented in this paper are in
atomic units ( $\hbar=m_e=e=1/4\pi\epsilon_0=1$).



\section{Methodology}

Since the properties of our interest involves atomic state functions (ASFs) 
of two-valence nature, we need an accurate multireference theory to calculate 
the many-body wavefunctions and the properties. Considering this, in the present work 
we have employed an all-particle FSRCC theory for two-valence \cite{mani-11b, ravi-21} systems 
to probe the clock transition and related properties in Sr.
In Ref. \cite{mani-17}, we have discussed in detail the implementation of 
FSRCC theory for closed-shell and one-valence systems to a sophisticated 
parallel code and have given the working equations and the contributing 
Goldstone diagrams. So, here, for completeness, we provide a very brief 
description of the FSRCC theory for two-valence systems and the properties 
calculations in the context of two-valence Sr.

\subsection{Two-valence FSRCC Theory}

\begin{figure}
 \includegraphics[scale=0.55, angle=0]{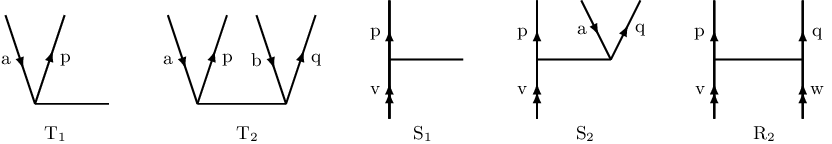}
 \caption{The diagrammatic representation of closed-shell, one-valence and 
	two-valence single and double CC operators.}
 \label{tsr-diag}
\end{figure}

The atomic state function for a two-valence atom or ion is obtained 
by solving the many-body Schrodinger equation
\begin{equation}
  H^{\rm DCB}|\Psi_{vw} \rangle = E_{vw} |\Psi_{vw} \rangle,
  \label{hdc_2v}
\end{equation}
where $|\Psi_{vw}\rangle$ is the exact many-body wavefunction and $E_{vw}$ 
is the corresponding exact energy. The indices $v, w, \cdots$ represent the 
valence orbitals. $H^{\rm DCB}$ is the Dirac-Coulomb-Breit no-virtual-pair 
Hamiltonian expressed as 
\begin{eqnarray}
   H^{\rm DCB} & = & \sum_{i=1}^N \left [c\bm{\alpha}_i \cdot
        \mathbf{p}_i + (\beta_i -1)c^2 - V_{N}(r_i) \right ]
                       \nonumber \\
    & & + \sum_{i<j}\left [ \frac{1}{r_{ij}}  + g^{\rm B}(r_{ij}) \right ],
  \label{ham_dcb}
\end{eqnarray}
where $\bm{\alpha}$ and $\beta$ are the Dirac matrices, and $1/r_{ij} $ 
and $g^{\rm B}(r_{ij})$ are the Coulomb and Breit interactions, respectively.
In FSRCC theory, $|\Psi_{vw} \rangle$ is written as
\begin{equation}
|\Psi_{vw}\rangle = e^T \left[ 1 + S_1 + S_2 + 
	\frac{1}{2} \left({S_1}^2 + {S_2}^2 \right) + 
	R_2 \right ]|\Phi_{vw}\rangle,
  \label{2v_exact}
\end{equation}
where $|\Phi_{vw}\rangle, = a^\dagger_wa^\dagger_v |\Phi_0\rangle,$
is the Dirac-Fock reference state for a two-valence system. Operators 
$T$, $S$ and $R$ are the electron excitation operators, referred to as 
the coupled-cluster (CC) operators, for closed-shell, one-valence 
and two-valence sectors, respectively. The subscripts $1$ and $2$ with 
these operators represent the single and double excitations, referred 
to as the coupled-cluster with singles and doubles (CCSD) approximation.
The FSRCC theory with CCSD approximation subsumes most of the electron 
correlation effects in atomic properties and provides an accurate description 
of the calculated properties. In the second quantized representation, 
these CC operators are expressed as
\begin{subequations}
\begin{eqnarray}
   T_1  = \sum_{ap}t_a^p a_p^{\dagger}a_a {\;\; \rm and \;\;} 
   T_2  = \frac{1}{2!}\sum_{abpq}t_{ab}^{pq} a_p^{\dagger}a_q^{\dagger}a_ba_a,
\end{eqnarray}
\begin{eqnarray}
   S_1 = \sum_{p}s_v^p a_p^{\dagger}a_v  {\;\; \rm and \;\;}
   S_2 = \sum_{apq}s_{va}^{pq} a_p^{\dagger}a_q^{\dagger}a_aa_v,
\end{eqnarray}
\begin{eqnarray}
  R_2 = \sum_{pq}r_{vw}^{pq} a_p^{\dagger}a_q^{\dagger}a_wa_v.
\end{eqnarray}
 \label{t1t2}
\end{subequations}
Here, the indices $a, b,\ldots$ and $p, q,\ldots$ represent the core and virtual 
orbitals, respectively. And, $t_{\ldots}^{\ldots}$, $s_{\ldots}^{\ldots}$ 
and $r_{\ldots}^{\ldots}$ are the cluster amplitudes corresponding to $T$, 
$S$ and $R$ cluster operators, respectively. The diagrammatic representations 
of these operators are shown in Fig. \ref{tsr-diag}.

The operators for closed-shell and one-valence sectors are obtained 
by solving the set of coupled nonlinear equations discussed in 
Refs. \cite{mani-09} and \cite{mani-10}, respectively. The two-valence 
CC operator, $R_2$, is obtained by solving the CC equation \cite{mani-11b, ravi-21}
\begin{eqnarray}
   \langle\Phi^{pq}_{vw}|
    \bar H_{\rm N} +
   \{\contraction{\bar}{H}{_{\rm N}}{S}\bar H_{\rm N}S^{'}\} +
   \{\contraction{\bar}{H}{_{\rm N}}{S}\bar H_{\rm N}R_2\}
   |\Phi_{vw}\rangle = \nonumber \\
   E^{\rm att}_{vw}
   \langle\Phi^{pq}_{vw}|\Bigl[S^{'} + R_2 \Bigr]|\Phi_{vw}\rangle.
   \label{ccsd_2v}
\end{eqnarray}
Here, for compact notation we have used $S'=S_1+S_2 + \frac{1}{2}(S^2_1 + S^2_2)$. 
$E^{\rm att}_{vw}$ is the two-electron attachment energy and is defined 
as the difference between the correlated energy of $(n-2)-$electron (closed-shell)
and $n-$electron (two-valence) sectors, $E_{vw} - E_0$. 
And, $\bar{H}_N, = e^{-T^{(0)}} H_N e^{T^{(0)}}$, is a similarity
transformed Hamiltonian, which using Wick's theorem, can be reduced
to the form
\begin{eqnarray}
  \bar{H}_N &=& H_N + \{\contraction[0.4ex]{} {H}{_N} {T} H_N T^{(0)}\} 
          + \frac{1}{2!} \{\contraction[0.4ex]{}{H}{_N}{T} 
         \contraction[0.8ex] {}{H}{_NT_0}{T^{(0)}}H_NT^{(0)}T^{(0)} \} 
                  +\nonumber \\ 
           && \frac{1}{3!}\{\contraction[0.4ex]{}{H}{_N}{T} 
          \contraction[0.8ex] {}{H}{_N T_0}{T^{(0)}}
          \contraction[1.2ex]{}{H}{_N T^{(0)} T_0}{T^{(0)}} 
          H_N T^{(0)} T^{(0)} T^{(0)} \} 
          + \frac{1}{4!}\{\contraction[0.4ex]{}{H}{_N}{T} 
          \contraction[0.8ex] {}{H}{_N T_0}{T^{(0)}}
          \contraction[1.2ex]{}{H}{_N T^{(0)} T_0}{T^{(0)}} 
          \contraction[1.6ex]{}{H}{_N T^{(0)}T^{(0)} T_0}{T^{(0)}} 
          H_N T^{(0)} T^{(0)} T^{(0)} T^{(0)} \}. 
\label{hnbar}   
\end{eqnarray}

\subsection{Perturbative triples correction to energy}

Considering the complicated nature of electron correlations in two-valence 
systems, it is important to include the correlation effects from triple 
excitations in FSRCC. In our previous work \cite{ravi-21}, we have implemented 
triples corrections to the properties of two-valence systems 
using perturbative approach. It should be noted that, the perturbative 
triples incorporate dominant contributions from triple excitations with 
far less computational cost than full triples. Considering this, in this work, 
we implement triples corrections to energy using perturbative
approach.

\begin{figure}
 \includegraphics[scale=0.45, angle=0]{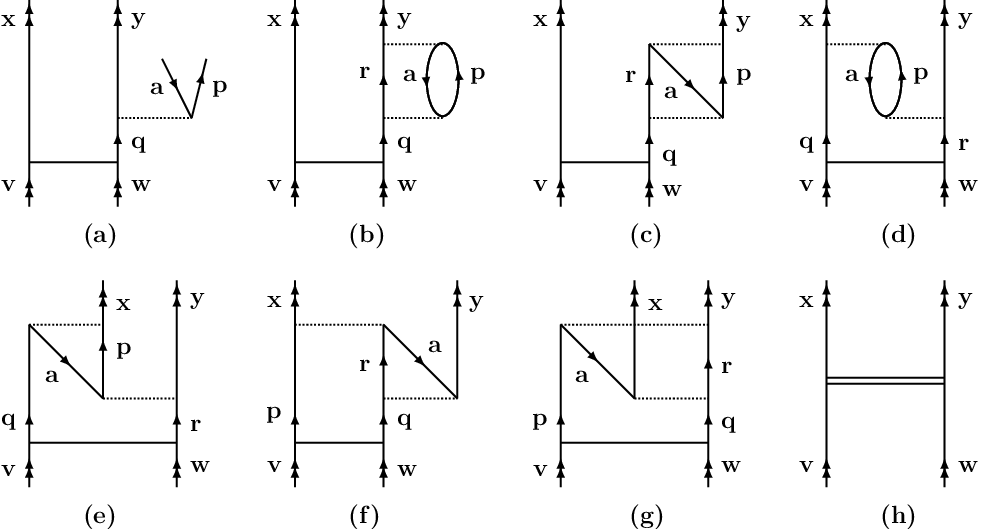}
	\caption{The Golstone diagrams contributing to perturbative 
	triples ${\widetilde R}_3$ (diagram (a)) and correlation 
	energy (diagrams (b - g)) arising from perturbative triples.
	Diagram (h) represents a two-body effective operator using 
	which valence-states are coupled to give atomic state 
	functions, as shown in Fig. \ref{coupling}.}
 \label{ptrb-diag}
\end{figure}

\begin{figure}
 \includegraphics[scale=0.48, angle=0]{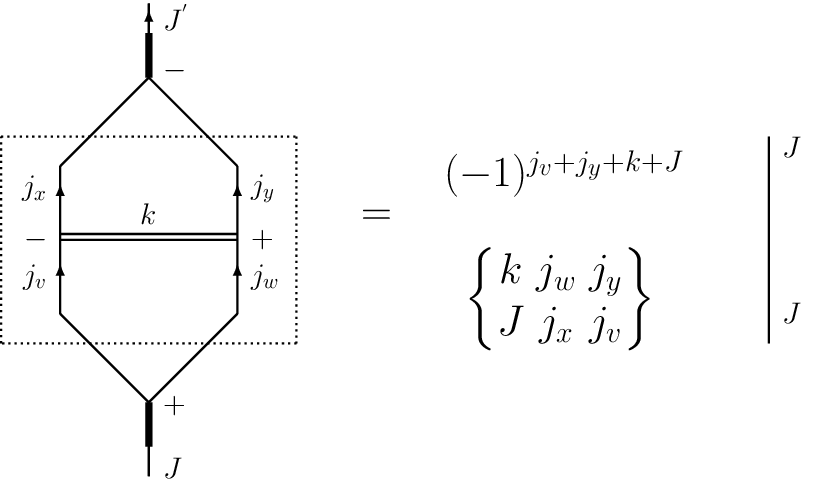}
 \caption{Angular factors arising from the coupling of single-particle 
	states of a two-body effective energy operator.}
 \label{coupling}
\end{figure}

Since magnitude of two-valence CC operator $R_2$ is larger than $T$ 
and $S$ operators for two-valence systems, for this, we choose the triples 
which arise from $R_2$. After a two-body residual interaction, 
$g_{ij} = \sum_{i<j}[\frac{1}{r_{ij}} + g^{\rm B}(r_{ij})]$, contracted 
with $R_2$, it leads to perturbative triples operator, 
${\widetilde R}_3 = \contraction[0.5ex]{}{g}{}{R} gR_2$.
The Goldstone diagram corresponding to ${\widetilde R}_3$ is shown in 
Fig. \ref{ptrb-diag}(a), which corresponds to the algebraic 
expression 
\begin{equation}
  \widetilde{R}_3 \approx \frac{1}{\Delta\epsilon^{xyp}_{vwa}}
  a^\dagger_x a^\dagger_y a^\dagger_p a_a a_w a_v
 \sum_{q}\langle yp|g|qa\rangle\langle xq|R_2|vw\rangle,
\end{equation}
where $\Delta\epsilon^{xyp}_{vwa} = \epsilon_{v}+\epsilon_{w}+\epsilon_{a}
-\epsilon_{x}-\epsilon_{y}-\epsilon_{p}$. The operator $\widetilde{R}_3$ 
now contract with residual Coulomb interaction to contribute to the energy. 
There are {\em six} energy diagrams from $\widetilde{R}_3$, shown as 
diagrams (b - g) in Fig. \ref{ptrb-diag}. These diagrams correspond to the 
algebraic expression
\begin{eqnarray}
  \{\contraction[0.6ex]{}{g}{}{\widetilde R} g {\widetilde R}_3\}^{xy}_{vw} &&= 
	\sum_{apqr}\frac{1}{\Delta\epsilon_{vwa}^{xrp}}
	\left [ \tilde{g}_{yarp}g_{rpqa}r_{xqvw} \right. \nonumber \\
&&\left. +\tilde{g}_{xaqp}g_{pyar}r_{qrvw} - g_{xapr}g_{ryqa}r_{pqvw} \right. \nonumber \\ 
&&\left. - g_{aypr}g_{xraq}r_{pqvw} \right ],
\end{eqnarray}
where $\tilde g_{ijkl} = g_{ijkl} - g_{jikl}$. The diagrams (b - g) 
are computed with respect to uncoupled states and stored in the form 
an effective diagram, digram (h). Next, we couple the single particle 
valence orbitals to give the atomic state function. Angular factors 
arising from this coupling is provided in Fig. \ref{coupling}.

\subsection{Hyperfine induced E1 transition rate}

Since Fermionic isotope of Sr possess non zero nuclear spin, the clock 
transition can be allowed through a hyperfine interaction. 
The hyperfine state, $|\Gamma F M_{F}\rangle$, is obtained by coupling an 
electronic state with the eigenstate of the nuclear spin operator ${\mathbf I}$. 
Considering the hyperfine interaction Hamiltonian, $H_{\rm HFS}$, as a 
perturbation, we can write, within the first-order time-independent 
perturbation theory,
\begin{equation}
    |\Gamma F M_{F}\rangle = \sum_{n} \left[\frac{\langle \gamma_{n} \Psi_{n} \gamma_{I}
    I| H_{HFS} | \gamma_{0} \Psi_{0} \gamma_{I} I \rangle}{E_{0} - E_{n}}\right]
    \times |\gamma_{n} \Psi_{n} \gamma_{I} I\rangle.
\end{equation}
The term within the bracket represents the hyperfine mixing of unperturbed 
state $|\gamma_{0} \Psi_{0} \gamma_{I} I\rangle$ with an excited 
state $|\gamma_{n} \Psi_{n} \gamma_{I} I\rangle$. The parameters $\Gamma$ 
and $\gamma_{i}$ are additional quantum numbers to identify the states 
uniquely, and $E$ is the exact energy.

The transition rate between two hyperfine levels $|\Gamma_i F_i M_{F_i} \rangle$ 
and $|\Gamma_f F_f M_{F_f} \rangle$ is expressed as \cite{johnsson_book}
\begin{eqnarray}
	\Gamma_{\rm HFS} = \frac{4 \alpha \omega^3}{3c^2} \frac{1}{(2F_i+1)}
	|E1_{\rm HFS}|^2,
\label{hfs-gamma}
\end{eqnarray}
where $\alpha$ is the fine structure constant, $c$ is the speed of light, 
and $\omega$ is the transition frequency. $E1_{\rm HFS}$ is referred 
to as the HFS induced electric dipole transition amplitude and, 
using the expressions for dipole and hyperfine matrix 
elements \cite{lindgren-86, johnsson_book}, can be derived to 
the form
\begin{eqnarray}
E1_{\text{HFS}} &=& c(I,J,F, \mu_I) 
\left[ \frac {\langle \Psi_f||\mathbf{t}^1||\bar \Psi_n \rangle \langle \bar \Psi_n||
	{\mathbf d}||\Psi_i \rangle}{E_{\bar n} - E_i} \right. \nonumber \\
&& \left. +\frac {\langle \Psi_f ||{\mathbf d}||{\Psi}_n\rangle\langle {\Psi}_n||
	\mathbf{t}^1||\Psi_i \rangle}{E_n - E_i} \right],
\label{e1hfs}
\end{eqnarray}
where c(I, J, F, $\mu_{I}$) is the angular factor associated with the 
coupling of nuclear and electronic angular states. Here, ${\mathbf d}$ 
is the dipole operator and ${\mathbf t}^1$ is the magnetic dipole 
hyperfine operator \cite{johnsson_book, brage_98}.
As evident from Eq. (\ref{e1hfs}), clock transition in Fermionic 
Sr can occur in two pathways. First, the initial state couples to the 
opposite parity intermediate states via a dipole transition, and 
then transitions to clock state via a hyperfine interaction. 
Alternatively, the initial state can couple to the same parity 
intermediate states through a hyperfine interaction and then connect 
to clock state through a dipole operator. For the present case 
of ${^{87}}$Sr: $|\Psi_i\rangle = 5s^2{\;^1}S_0$, 
$|\Psi_f\rangle = 5s5p{\;^3}P^{o}_0$, $|\bar \Psi_n\rangle = 5s5p{\;^3}P^{o}_1$, 
$5s5p{\;^1}P^{o}_1$, $5s6p{\;^3}P^{o}_1$, $5s6p{\;^1}P^{o}_1$, and 
$|\Psi_n\rangle = 5s6s{\;^3}S_1$, $5s4d{\;^3}D_1$.

\begin{figure}
\includegraphics[scale=0.3, angle=0]{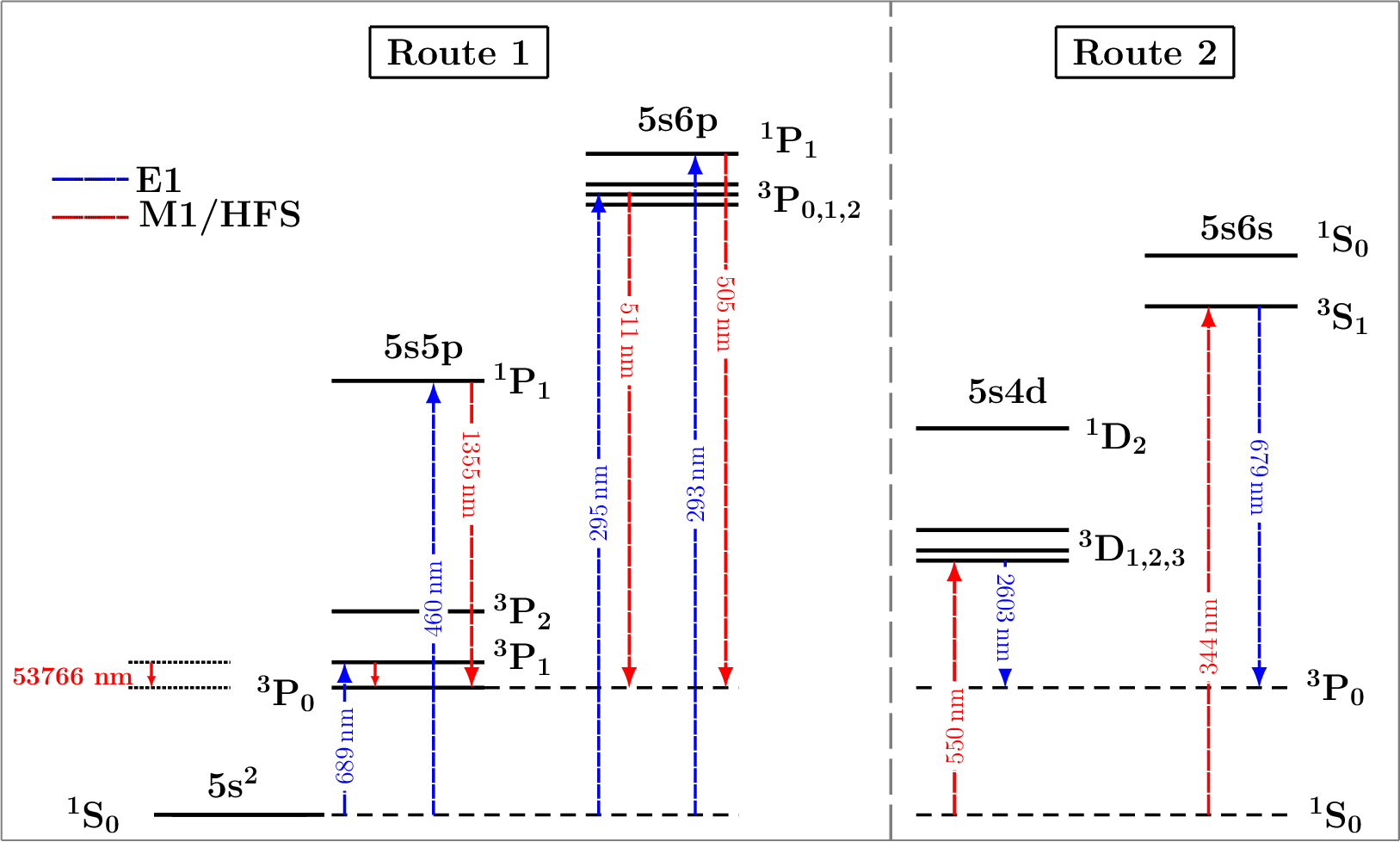}
\caption{Schematic energy level diagram showing the $5s^2{\;^1}S_0 - 5s5p{\;^3P^o_0}$ 
	clock transition in $^{87}$Sr and $^{88}$Sr via a HFS induced and 
	a two-photon E1+M1 channel, respectively.}
 \label{fig_e1m1}
\end{figure}

\subsection{E1M1 transition rate}

In Bosonic isotopes of Sr, since they possess a zero nuclear spin, the clock 
transition can occur via a {\em {two-photon}} E1 + M1 transition. Like the 
case of Fermionic Sr, it can happen in two pathways. In the first, the initial 
state $|\Psi_i \rangle$ couples to opposite parity intermediate states $|\bar 
\Psi_n\rangle$ via an electric dipole transition, and then decays to final state $|\Psi_f 
\rangle$ through a magnetic dipole transition (emitting a photon of energy $\omega_1$). 
And in the second pathway, the initial state couples with same parity
intermediate states $|\Psi_n\rangle$ via a magnetic dipole operator (emitting a 
photon of energy $\omega_2$), then connects to final state through an 
electric dipole transition. Mathematically, the E1+M1 transition rate 
between $|\Psi_i \rangle$ and $|\Psi_f \rangle$ can be expressed in 
terms of the reduced matrix elements of the E1 and M1 
operators, as \cite{santra-04, palki-24}
\begin{equation}
  \Gamma_{E1M1} = \frac{8}{27\pi}\alpha^6 \int_{0}^{\infty}d\omega_{1}
	\omega_{1}^{3}\int_{0}^{\infty}d\omega_{2}\omega_{2}^{3} \; |E1_{\rm M1}|^2 ,
  \label{e1m1-gamma}
\end{equation}
where E1+M1 transiton amplitude is expressed as
\begin{eqnarray}
	E1_{\rm M1} && = \left[  \frac {\langle \Psi_f||{m}_{1}||\bar \Psi_n\rangle \langle 
	     \bar \Psi_n||{\mathbf d}||\Psi_i \rangle} {E_{\bar n} + \omega_{1} - E_{i}} \right.  \nonumber \\
	&& \left. + \frac {\langle \Psi_f||{\mathbf d}|| \Psi_n\rangle \langle 
	      \Psi_n||m_1||\Psi_i \rangle} {E_n + \omega_{2} - E_{i}} \right] \nonumber \\
	&& \times \delta (E_{i} + \omega_{1} + \omega_{2} - E_{f}).
\label{e1m1-gamma2}
\end{eqnarray}
Here, since the transition is allowed through a two photon channel, 
the energy difference of final and initial states satisfies 
$E_f - E_i = \omega_1 + \omega_2$.

The reduced matrix elements in Eqs. (\ref{e1hfs}) and (\ref{e1m1-gamma2}) 
are calculated using FSRCC theory. The details related to the calculation
is provided in our previous works \cite{mani-11b, ravi-21}.

\subsection{Isotope Shift}

Isotope shifts in energy of an atom or ion arise from the variations 
in nuclear mass and charge distributions between two isotopes. 
Incorporating isotope shifts in the calculations of atomic properties 
has become important due to the pressing need to match atomic theory 
calculations with high accuracy atomic experiments. 
The isotope shifts in energy of a two-valence atom or ion within the 
framework of multireference coupled-cluster theory can be calculated as 
\begin{equation}
  \Delta E_{\rm iso} = \frac{\langle \Psi_{vw} |H_{\rm iso}| \Psi_{vw} \rangle}
  {\langle \Psi_{vw}| \Psi_{vw}\rangle},
  \label{iso_eqn}
\end{equation}
where $|\Psi_{vw}\rangle$ is an exact atomic state function obtained 
by solving Eq. (\ref{hdc_2v}) using multireference FSRCC theory, and 
operator $H_{\rm iso}$ is an isotope shift Hamiltonian. $H_{\rm iso}$ 
embeds two important contributions: the field shift ($H_{\rm FS}$), 
which arises due to varying nuclear charge distribution between two 
isotopes; and the mass-shift ($H_{\rm MS}$), which occurs 
due to recoil motion of the nucleus. The mass-shift Hamiltonian can 
further be separated to have two contributions, $H_{\rm MS} = H_{\rm NMS} + H_{\rm SMS}$, 
where $H_{\rm NMS}$ and $H_{\rm SMS}$ are referred to as the normal mass-shift 
and specific mass shift operators, respectively. The $H_{\rm NMS}$ 
is a one-body operator, and expressed as \cite{naze-13, gaidamauskas_11}
\begin{eqnarray}
	H_{\rm NMS}  = \frac{1}{2M}\sum_{i=1}^{N} && 
	\left[ {\mathbf p}^2_i - \frac{\alpha Z}{r_i} \big( ({\mathbf \alpha}_i \cdot {\mathbf p}_i)
                                           \right. \nonumber \\
	&& \left. - ({\mathbf \alpha}_i \cdot {\mathbf C}^1_i) ({\mathbf C}^1_i \cdot {\mathbf p}_i)  \big) \right],
\end{eqnarray}
where $M$ is the nuclear mass, ${\mathbf \alpha}_i$ is the Dirac 
matrix, $\alpha$ is the fine structure
constant, $Z$ is the atomic number, and ${\mathbf p}_i$ and ${\mathbf C^1}$ 
are the momentum and Racah operators, respectively. The specific mass-shift 
operator $H_{\rm SMS}$, on the other hand, is a two-body operator, and 
expressed as \cite{naze-13, gaidamauskas_11}
\begin{eqnarray}
H_{\rm SMS} = \frac{1}{M}\sum_{i < j}^{N} &&
  \left [ {\mathbf p_i}\cdot {\mathbf p_j} - \frac{\alpha Z}{r_i} \big( ({\mathbf \alpha}_i \cdot {\mathbf p}_j)
                                                 \right. \nonumber \\
  && \left. - ({\mathbf \alpha}_i \cdot {\mathbf C}^1_i) ({\mathbf C}^1_i \cdot {\mathbf p}_j)  \big) \right],
\end{eqnarray}
Considering the nucleus as a uniformly charged sphere of radius R, the field-shift Hamiltonian 
$H_{\rm FS}$ represents the modification in the nuclear potential experienced by an electron
due to a small change $\delta R$ in the nuclear charge radius. This perturbation accounts for
the variation in the electronic energy arising from differences in the nuclear charge distribution 
between isotopes and, therefore, is responsible for the field (or volume) contribution to the
isotope shift. The corresponding Hamiltonian can be expressed \cite{johnsson_book}
\begin{eqnarray}
 H_{\rm FS} = 
 \left\{ \begin{array}{c}
	 - \frac{5Z}{4R^3} \left[1 - r^2/R^2 \right] {\rm \;for\;} r \le R \\
	 0                                       {\rm \;for\;} r > R.
\end{array} \right.
\end{eqnarray}

The details, such as one- and two-body operators' matrix elements, contributing 
Goldstone diagrams to Eq. (\ref{iso_eqn}), algebraic expressions and angular 
factors, etc., related to the implementation of isotope shifts calculations 
for one- and two-valence systems within the framework of FSRCC theory shall 
be provided in a separate work \cite{palki-25}.

\section{Results and discussions}
\label{results}
 
\subsection{Single-particle basis and convergence of properties}

Accurate single-particle basis is essential to get reliable results using 
FSRCC and PRCC theories. In this work, we have employed Gaussian-type 
orbitals (GTOs) \cite{mohanty-91} as the single-electron basis. GTOs 
form a finite basis set where single-particle wavefunctions are 
expressed as linear combinations 
of Gaussian-type functions (GTFs). The large component of 
the wavefunction is expressed as
\begin{equation}
  g^L_{\kappa p} (r) = C^L_{\kappa i} r^{n_\kappa} e^{-\alpha_p r^2},
\end{equation}
where $p = 0$, $1$, $2$, $\ldots$, $N$ refers to the GTO index and 
$N$ is the total number of GTFs. The exponent $\alpha_p$ is further 
expressed as $\alpha_0 \beta^{p-1}$, where $\alpha_{0}$ and $\beta$ 
are two independent parameters. Parameters $\alpha_{0}$ and $\beta$ 
are optimized separately for each orbital symmetry to ensure that 
the single-electron wavefunctions and energies match well with 
the numerical values obtained from GRASP2K \cite{jonsson-13}. 
The small components of the wavefunctions are derived from 
the large components by applying the kinetic balance condition \cite{stanton-84}. 

\begin{table}
\begin{center}
\caption{Comparison of single-particle and SCF energies (in hartree) from GTO with
        GRASP2K and B-Spline results. The optimized parameters, $\alpha_0$ and $\beta$, 
        for the even-tempered basis used in our calculations are also included. }
\begin{ruledtabular}
\begin{tabular}{cccc}
  Orbitals  &  GTO   & GRASP2K  & B-Spline \\
\hline
$1s_{1/2}$  &  $-596.12431$   &  $-596.12414$     &  $-596.12464$   \\
$2s_{1/2}$  &   $-83.64336$   &   $-83.64329$     &   $-83.64334$   \\
$2p_{1/2}$  &   $-75.87558$   &   $-75.87551$     &   $-75.87553$   \\ 
$2p_{3/2}$  &   $-73.34196$   &   $-73.34194$     &   $-73.34194$   \\      
$3s_{1/2}$  &   $-14.45568$   &   $-14.45567$     &   $-14.45568$   \\ 
$3p_{1/2}$  &   $-11.57593$   &   $-11.57592$     &   $-11.57592$   \\
$3p_{3/2}$  &   $-11.17082$   &   $-11.17081$     &   $-11.17082$   \\
$3d_{3/2}$  &    $-6.12644$   &    $-6.12645$     &    $-6.12643$   \\
$3d_{5/2}$  &    $-6.05602$   &    $-6.05602$     &    $-6.05601$   \\ 
$4s_{1/2}$  &    $-2.43444$   &    $-2.43445$     &    $-2.43444$   \\
$4p_{1/2}$  &    $-1.61374$   &    $-1.61374$     &    $-1.61374$   \\ 	       
$4p_{3/2}$  &    $-1.56683$   &    $-1.56683$     &    $-1.56683$   \\ 
\hline
	$E_{\rm SCF}$ &  $-3177.52169$  &    $-3177.52166$     &  $-3177.52251$ \\
\hline
  Parameter  &  $s$   & $p$  &  $d$        \\
\hline
  $\alpha_0$ & 0.00539 & 0.00567 & 0.00494 \\
  $\beta$    & 1.995   & 1.982   &  2.045  \\
 \end{tabular}
\end{ruledtabular}
\label{grasp-en}
\end{center}
\end{table}

\begin{figure*}
  \includegraphics[scale=0.35, angle=-90]{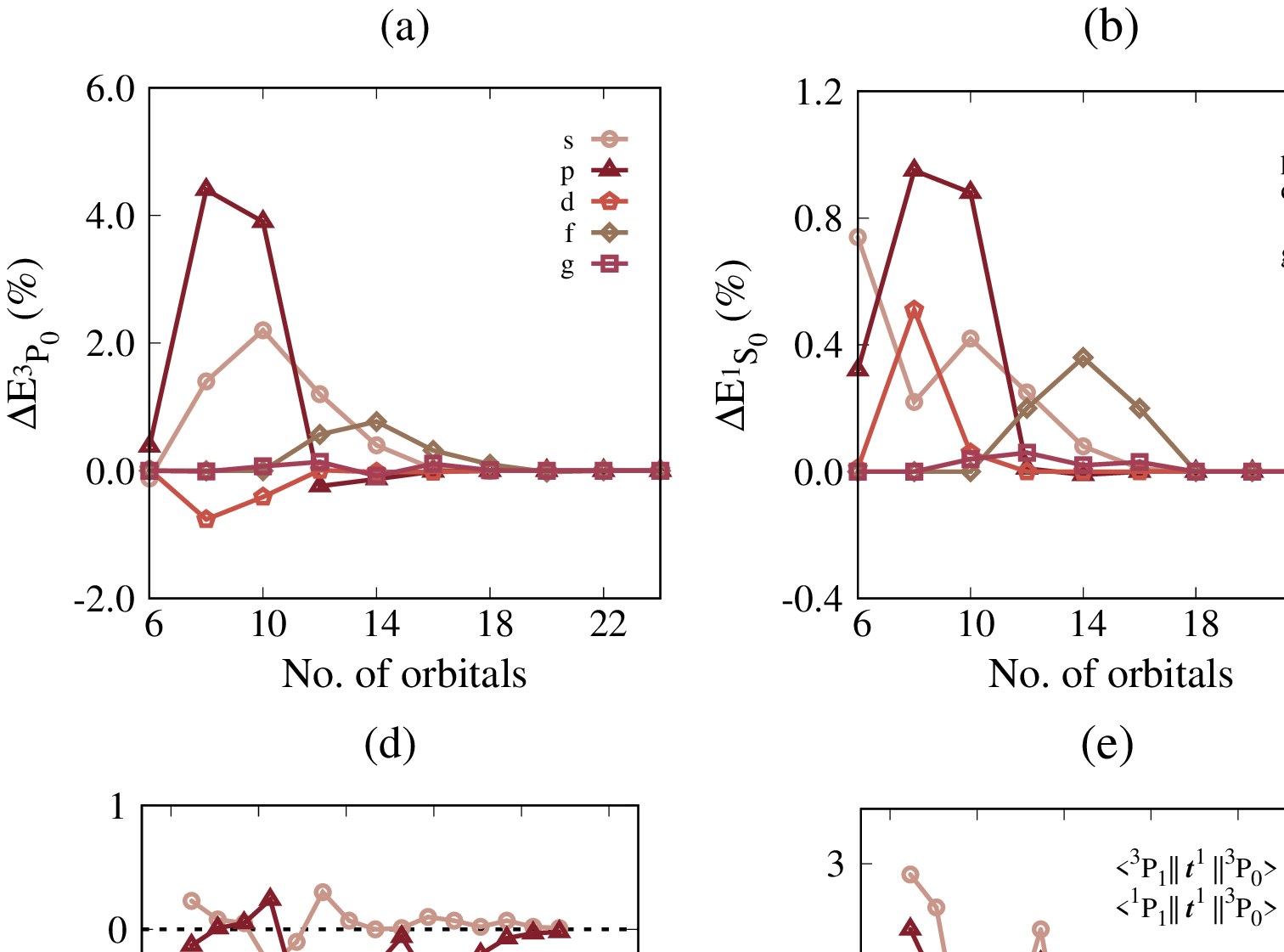}
	\caption{Panels (a - c) show the convergence trend for energy 
	of $^3P_{0}$, $^1S_{0}$ and $^1P_{1}$ states with augmentation 
	of orbitals in $s$, $p$, $d$, $f$ and $g$ symmetries.
        Panels (d - f) show the convergence trend for E1, HFS, and M1 
	reduced matrix elements with basis size.}
 \label{conv-diag}
\end{figure*}

In Table \ref{grasp-en}, we list the values of optimized $\alpha_{0}$ and
$\beta$ parameters along with the single-particle and 
self-consistent-field (SCF) energies for Sr. For comparison, we have 
also provided the results from GRASP2K \cite{jonsson-13} and 
B-spline \cite{zatsarinny-16} calculations. The single-electron basis 
used in our calculations also incorporate the effects of vacuum polarization 
and self-energy corrections. As evident from the table, both 
single-particle as well as SCF energies from GTO show an excellent 
agreement with GRASP2K and B-spline results. The maximum differences 
are observed to be 0.00003\% and 0.00006\% for SCF and single-particle 
energies, respectively.

Since GTOs form a mathematically incomplete basis, it is essential 
to check the convergence of the properties with basis size. 
Panels (a - c) of Fig. \ref{conv-diag} show the percentage change in the excitation
energies for 
$^3P_{0}$, $^1S_{0}$ and $^1P_{1}$ states, respectively, as a function of 
number of orbitals in $s$, $p$, $d$, $f$, and $g$ symmetries. 
As can be expected, lower energy orbitals in each symmetry show a 
stronger electron correlation effect. The contribution to energy 
decreases with an increase in the number of orbitals in each symmetry. 
For $^3P_{0}$ and$^1S_{0}$ states, the dominant electron correlation 
is observed from $s$, $p$, $d$ and $f$-electrons, whereas for $^1P_{1}$ state, 
it is from $s$, $p$ and $f$-electrons.
From a systematic analysis of calculations, we find that the energies 
converge well with $19s$, $17p$, $17d$, $14f$, and $8g$ orbitals, 
yielding a total basis size of 131. 
We also checked the contributions from the orbitals from $h$-symmetry 
and higher, however, their contributions to energy were found to be negligible.
Similarly, to check the convergence of the properties, we calculated E1, 
HFS and M1 matrix elements as a function of basis size. As evident from 
the Table \ref{conv_tab}, we begin with a moderate size basis of 68 
orbitals and systematically add orbitals in each symmetry until the 
change in the property becomes negligible. Panels (d - f) show the 
percentage change as a function of basis size for some key E1, HFS 
and M1 matrix elements as an example. As discernible from the figures, all the
three properties converge well with basis size. Other key point to 
note is that, properties require larger basis to converge than energy.

\begin{table*}
\caption{Two-electron removal energy for $^1S_0$ state (cm$^{-1}$) 
	and excitation energies for some low-lying even- and odd-parity 
	excited states of Sr. The results listed correspond to $5s^2$, 
	$5s5p$, $5s4d$, $5s6s$, $5s6p$ configurations in the model space. 
	For quantitative analysis of electron correlations, contributions 
	from the Breit interaction, QED corrections and perturbative 
	triples are listed separately.}
\begin{ruledtabular}
\begin{tabular}{cccccccccc}
\multicolumn{1}{c}{\textrm{States}}  &
\multicolumn{1}{c}{\text{DC-CCSD}}&
\multicolumn{1}{c}{\text{Breit}}&
\multicolumn{1}{c}{\text{Self-energy}}&
\multicolumn{1}{c}{\text{Vac-pol}} &
\multicolumn{1}{c}{\text{Triples}} &
\multicolumn{1}{c}{\text{Total}}&
\multicolumn{1}{c}{\text{Other cal.}} & 
\multicolumn{1}{c}{\text{NIST\cite{NIST-ASD}}} &     
\multicolumn{1}{c}{\text{\% Error}}  \\                                  
\hline     
$5s^{2}$ $ ^1S_0$  &135785 & 0 & 0 & 4 & 238 & 136027 & 136244$^{b}$, 135444$^{c}$, 135940$^{f}$ & 134897$^{a}$ & 0.84  \\              
$5s5p$\ $ ^3P_0$   &14901  & 4 & 0 & 6 & $-393$ & 14518 &  14241$^{d}$, 14806$^{b}$, 14550$^{c}$  & 14318$^{a}$  &  1.4 \\
                   &       &   &   &   &        &       &  12490$^{e}$, 14860$^{f}$               &              &        \\           
$5s5p$\ $ ^3P_1$   &15176  & 3 & 0 & 6 & $-387$ & 14798 & 14995$^{b}$, 14739$^{c}$, 14448$^{d}$   & 14504$^{a}$  &  2  \\
                   &       &   &   &   &        &       & 12663$^{e}$, 14854$^{f}$                &              &          \\         
$5s5p$\ $ ^3P_2$   &15720  & 0 & 0 & 6 & $-382$ & 15344 &  15399$^{b}$, 15142$^{c}$, 14825$^{d}$  & 14899$^{a}$  &  2.9    \\ 
                   &       &   &   &   &        &       &  13022$^{e}$, 14783$^{f}$               &              &            \\      
$5s4d$\ $ ^3D_1$   & 19840 & $-10$ & 0 & $-25$  & $-1719$ & 18086 & 18255$^{b}$, 18327$^{c}$, 18076$^{d}$ & 18159$^{a}$ & $-0.4$ \\
                   &       &       &   &        &         &       & 19571$^{e}$, 18409$^{f}$              &             &      \\   
$5s4d$\ $ ^3D_2$   & 19859 & $-10$ & 0 & $-29$  & $-1722$ & 18098 & 18298$^{b}$, 18394$^{c}$, 18141$^{d}$ & 18218$^{a}$ & $-0.6$ \\
                   &       &       &   &        &         &       & 19587$^{e}$, 18569$^{f}$              &             &         \\
$5s4d$\ $ ^3D_3$   & 19906 & $-12$ & 0 & $-37$  & $-1719$ & 18138 & 18422$^{b}$, 18506$^{c}$, 18254$^{d}$ & 18319$^{a}$ & $-0.9$ \\  
                   &       &       &   &        &         &       & 19617$^{e}$, 18747$^{f}$              &             &         \\   
$5s4d$\ $ ^1D_2$   & 22026 & $-9$  & 0 & $-24$  & $-2406$ & 19587 & 20428$^{b}$, 20441$^{c}$, 19968$^{d}$ & 20149$^{a}$ & $-2.8$ \\
                   &       &       &   &        &         &       & 20166$^{e}$, 20509$^{f}$              &             &         \\   
$5s5p$\ $ ^1P_1$   & 22834 &  1    & 0 &   5    &  $-505$ & 22335 & 21955$^{b}$, 21823$^{c}$, 21469$^{d}$ & 21698$^{a}$ &  2.9  \\ 
                   &       &       &   &        &         &       &  20832$^{e}$, 23387$^{f}$             &             &        \\
$5s6s$\ $ ^3S_1$   & 30375 &  1    & 0 &   5    &  $-242$ & 30139 & 29369$^{b}$, 29223$^{c}$              & 29038$^{a}$ & 3.8 \\ 
                   &       &       &   &        &         &       & 29019$^{d}$, 27488$^{e}$              &             &     \\
$5s6s$\ $ ^1S_0$   & 32292 &  1    & 0 &   5    & $-1045$ & 31253 & 30938$^{b}$, 30777$^{c}$              & 30591$^{a}$ & 2.2 \\
$5s6p$\ $ ^3P_0$   & 35747 &  1    & 0 &   6    &  $-321$ & 35433 & 34241$^{b}$, 34055$^{c}$              & 33853$^{a}$ & 4.7 \\
$5s6p$\ $ ^3P_1$   & 35803 &  1    & 0 &   6    &  $-791$ & 35019 & 34255$^{b}$, 34071$^{c}$              & 33868$^{a}$ & 3.4 \\
                   &       &       &   &        &         &       & 33814$^{d}$, 32110$^{e}$              &             &      \\
$5s6p$\ $ ^3P_2$   & 35939 &  0    & 0 &   6    &  $-316$ & 35629 & 34365$^{b}$, 34134$^{c}$              & 33973$^{a}$ & 4.9 \\
$5s6p$\ $ ^1P_1$   & 36256 &  0    & 0 &   7    &  $-750$ & 35513 & 34476$^{b}$, 34308$^{c}$, 34105$^{d}$ & 34098$^{a}$ & 4.1 \\
                   &       &       &   &        &         &       & 32487$^{e}$                           &             &      \\
\end{tabular}
\end{ruledtabular}
\begin{flushleft}
 $^{\rm a}$  Ref. \cite{NIST-ASD} - Expt. , 
 $^{\rm b}$ Ref. \cite{safronova-13} - CI + MBPT, 
 $^{\rm c}$ Ref. \cite{safronova-13} - CI + all-order, 
 $^{\rm d}$ Ref. \cite{porsev-08} - CI + MBPT , 
 $^{\rm e}$ Ref. \cite{porsev-08} - CI,
 $^{\rm f}$ Ref. \cite{mani-11} - FSRCC.
\end{flushleft}
\label{large-ee}
\end{table*}

\begin{figure}
\includegraphics[scale=0.30, angle=-90]{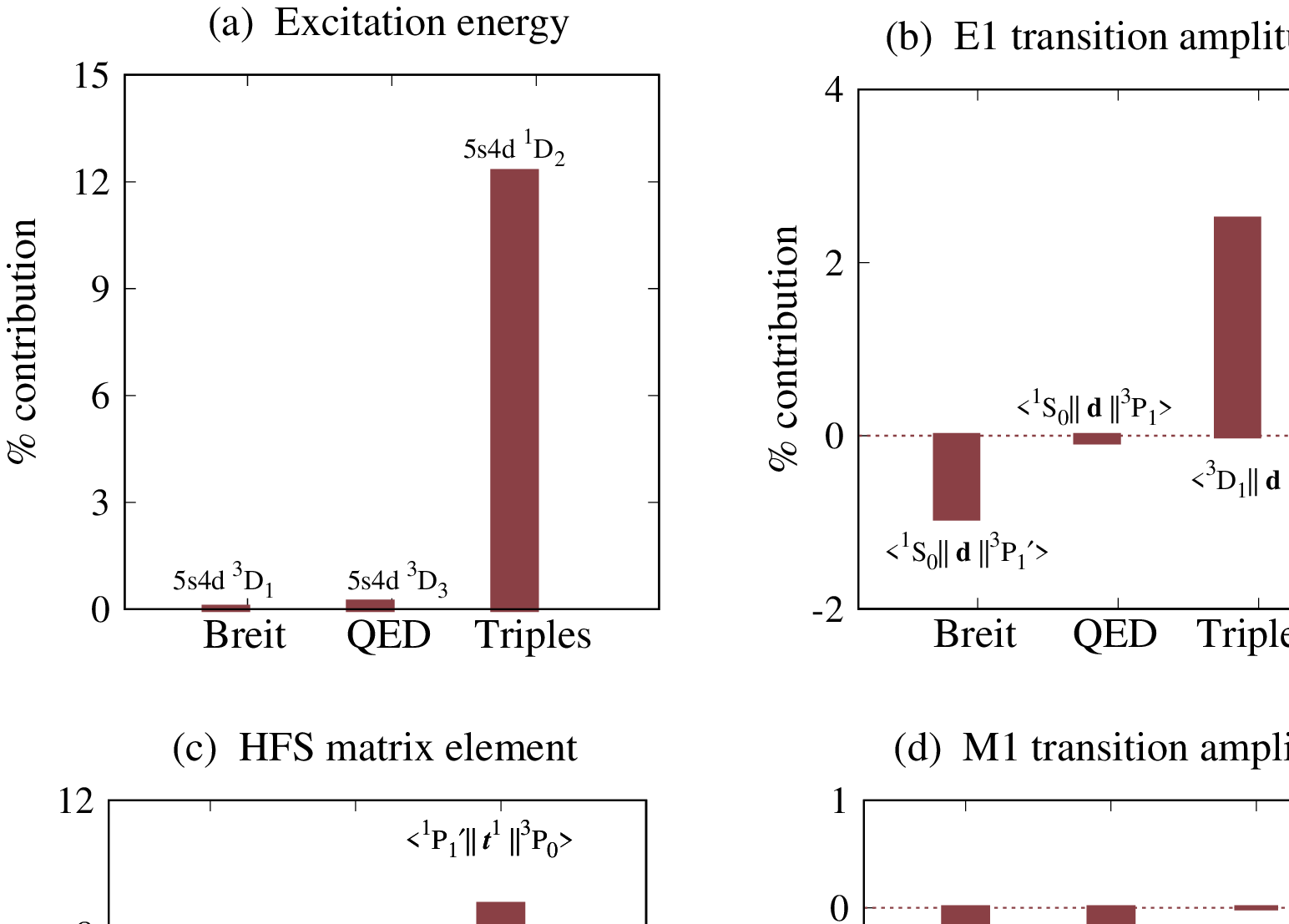}
\caption{Dominant percentage contributions from the Breit interaction, 
	QED effects and perturbative triples to excitation energy (panel (a)), 
	E1 transition amplitudes (panel (b)), HFS matrix elements (panel (c)), 
	and M1 transition amplitudes (panel (d)) for Sr.}
 \label{tbq-diag}
\end{figure}

\subsection{Excitation Energy}

In Table \ref{large-ee}, we present the excitation energies for several
low lying states of Sr from FSRCC calculation. The excitation energy 
of a general state $nln'l'{\;^{(2S+1)}}L_J$ can be calculated as
\begin{equation}
     \Delta E_{nln^{'}l^{'}{\;^{(2S+1)}}L_J} =
     E_{nln^{'}l^{'}{\;^{(2S+1)}}L_J} - E_{ns^2{\;^1}S_0},
\end{equation}
where $E_{ns^2{\;^1}S_0}$ and $E_{nln^{'}l^{'}{\;^{(2S+1)}}L_J}$ are the 
exact energies of the ground and excited states, respectively.
In Table \ref{large-ee}, we have also provided the experimental
energies and the results from other theory calculations for comparison. 
To improve the accuracy of our energies, we have incorporated the 
corrections the from Breit interaction and QED effects in our calculations.  
The other two important points to note in our study are, the inclusion 
of high energy configurations in the model space and the corrections 
from the perturbative triples. High energy configurations are essential 
to incorporate in multireference systems to account for {\em valence-valence} 
electron correlation accurately. In this work, we include $5s4d$, $5s6s$, 
and $5s6p$ high energy configurations in model space, in addition to the 
bare even and odd parity configurations $5s^2$ and $5s5p$, respectively. 
It is to be mentioned that, while two-valence FSRCC calculations with 
extended model spaces are quite challenging due to {\em intruder} states 
problem, we observed significant reductions to the errors in the energies. 
Similarly, perturbative triples are observed improve the energies 
significantly. Fig. \ref{tbq-diag}(a) shows the largest percentage 
contributions from Breit, QED and perturbative triples corrections 
to the energy. The largest contribution from perturbative triples is 
found to be $\approx$ 12\% in the case of 5s4d $^1D_{2}$ state. Whereas, 
the Breit and cumulative QED corrections are observed to $\approx$ 0.1 
and 0.2\%, respectively, in the case of 5s4d$^3D_{1}$ and 5s4d$^3D_{3}$ states.

Table \ref{large-ee} also shows the relative errors in our excitation energies 
with respect to NIST data. As evident, despite our 
efforts to improve energies by incorporating different corrections and 
high energy configurations, we get a large variation in the errors. 
The smallest and largest errors in our calculations 
are 0.4 and 4.9\%, respectively, in the case of 5s4d$^3D_{1}$ and 
5s6p$^3P_{2}$ states. Unlike the case of one-valence atoms or ions, large errors 
in the {\em ab initio} calculations of multireference systems are anticipated 
due to the complicated nature of electron correlations. Typically, there are three 
types of electron correlations exhibited by a multireference atom or ion; 
{\em core-core}, {\em core-valence} and {\em valence-valence}. It is extremely 
challenging to treat all these three accurately to the same level within 
the framework of a single theory. In the present work, within the framework 
of a FSRCC theory, {\em core-core} and {\em core-valence} correlations are
accounted to all orders of residual Coulomb interaction using a nonlinear 
RCC theory. And, the {\em valence-valence} correlation is treated in the 
framework of CI through an effective Hamiltonian approach. 
Looking into the literature, to the best of our search, we could find 
three previous works which report excitation energies for two-valence 
Sr \cite{safronova-13, porsev-08, mani-11}. 
Ref. \cite{safronova-13} uses CI+MBPT and CI+all-order method, whereas 
Ref. \cite{porsev-08} has employed CI+MBPT and CI methods to compute excitation 
energies. The third calculation \cite{mani-11} is using the same method 
as the present work, however, with some key additions in the present work.
Consistent with the trends in our calculations, the previous calculations 
also show large errors. For example, consider the case of metastable clock 
state 5s5p$^3P_{0}$, which is of the primary interest of our work, the errors 
in excitation energy are $\approx$ 3.4\% CI+MBPT \cite{safronova-13}, 1.6\% 
CI+all-order \cite{safronova-13}, 0.5\% CI+MBPT \cite{porsev-08}, 
12.8\% CI \cite{porsev-08}, and 3.8\% FSRCC \cite{mani-11}. The error 
in the present calculation is $\approx$ 1.4\%. The reason for the difference 
in the excitation energy from Ref. \cite{mani-11} could be attributed to 
the inclusion of the contributions from {\em nonlinear} terms in FSRCC,
high energy configurations, and Breit and QED corrections in 
the present work.

\subsection{E1, HFS and M1 Reduced Matrix Elements}

In Table \ref{tab-e1m1}, we present our results on E1, HFS and M1 reduced 
matrix elements from FSRCC calculations. These reduced matrix elements, 
along with the energies, are required to calculate the lifetime of the
metastable clock state. To improve the accuracy of these matrix elements,
we have incorporated the corrections from relativistic and QED effects, 
and also from the perturbative triples in our calculations. It is to be 
also mentioned that, the results presented in the table correspond 
to a larger model space with $5s^2$, $5s5p$, $5s4d$, $5s6s$, $5s6p$ 
configurations. 

\begin{table*}
\caption{E1, HFS and M1 reduced matrix elements (in a.u.) for Sr with
    $5s^2$, $5s5p$, $5s4d$, $5s6s$, $5s6p$ configurations in the
    model space. The data from experiments and other theory calculations 
	are also provided for comparison (only magnitude).}
\begin{ruledtabular}
\begin{tabular}{ccccccccclr}
\multicolumn{1}{c}{\textrm{States}}  &
\multicolumn{1}{c}{\textrm{DF}}  &
\multicolumn{1}{c}{\text{DC-CCSD}}&
\multicolumn{1}{c}{\text{Breit}}&
\multicolumn{1}{c}{\text{QED}}&
\multicolumn{1}{c}{\text{Breit + QED}}&
\multicolumn{1}{c}{\text{Triples}}&
\multicolumn{1}{c}{\text{Total}}&
\multicolumn{1}{c}{\text{Other cal.}}&
\multicolumn{1}{c}{\text{Expt.}} \\
\hline
\multicolumn{10}{c}{E1 Reduced Matrix Elements} \\
$\langle^1S_0||{\mathbf d}||^3P_1\rangle$ & 0.1325  & 0.1404  & $-0.0011$ & $-0.0001$  & $-0.0012$ & 0.0025 & 0.1417 & 0.160$^{a}$, 0.158$^{b}$ & 0.1555$^{c}$, 0.1510$^{d}$ \\
                                  &   &  &   &  &  &  &    & 0.13$^{e}$, 0.151$^{f}$  & 0.1486$^{g}$, 0.151$^{h}$    \\
                                  &   &  &   &  &  &  &    & 0.1311$^{i}$, 0.16$^{j}$ & 0.146$^{k}$                   \\
                                  &   &  &   &  &  &  &    & 0.152$^{l}$              &                                \\
$\langle^1S_0||{\mathbf d}||^1P_1\rangle$ & 4.6694 & 4.8934 & 0.0 & $-0.0005$ & $-0.0005$ & $-0.0124$ & 4.8805 & 5.28$^{a}$, 5.272$^{b}$  & 5.57$^{t}$, 5.40$^{k}$\\
                                  &   &  &  &  &   &  &  &   5.15$^{e}$, 5.248$^{f}$  & 5.248$^{h}$, 5.269$^{m}$    \\
                                  &   &  &  &  &   &  &  &   5.298$^{n}$, 5.673$^{o}$ &     \\
                                  &   &  &  &  &   &  &  &   5.367$^{i}$, 5.292$^{p}$ &     \\
                                  &   &  &  &  &   &  &  &   5.307$^{l}$              &     \\
$\langle^1S_0||{\mathbf d}||^3P^\prime_1\rangle$ & 0.1308    & 0.3384    & $-0.0033$ & $-0.0002$ & $-0.0035$ & $-0.0025$ & 0.3324  & & \\
$\langle^1S_0||{\mathbf d}||^1P^\prime_1\rangle$ & $-1.2695$ & $-2.6213$ & $-0.0005$ & $-0.0001$ & $-0.0006$ &    0.0314 & $-2.5905$ & 0.281$^{b}$, 0.236$^{a}$ &  0.26$^{k}$\\
$\langle^3S_1||{\mathbf d}||^3P_0\rangle$  & 2.1231 & 1.9686 & 0.0 & 0.0002 & 0.0002  & 0.0046 & 1.9734 & 1.96$^{a}$, 1.962$^{b}$ & 2.03$^{u}$\\
$\langle^3D_1||{\mathbf d}||^3P_0\rangle$  & $-2.6236$  & $-2.4329$ & $-0.0002$ & 0.0001 & $-0.0001$  & $-0.0627$ & $-2.4957$ & 2.74$^{a}$, 2.675$^{b}$ & 2.5$^{v}$, 2.7$^{w}$\\ \\
\hline
\multicolumn{10}{c}{HFS Reduced Matrix Elements ($\times$ 10$^{-7}$)} \\
$\langle^3P_1||t^1||^3P_0\rangle$ & 2.7867 & 2.8449 & $-0.0006$ & 0.0046  & 0.0040   & 0.0290 & 2.8779 & $2.023^{r}$ & \\
$\langle^1P_1||t^1||^3P_0\rangle$ & 1.9675 & 2.0534 &   0.0003  & 0.0030  & 0.0033   & 0.0356 & 2.0923 & $1.450^{r}$ & \\
$\langle^3P^\prime_1||t^1||^3P_0\rangle$ & 0.0122 & $-0.7961$ & $-0.0004$ & 0.0001  & $-0.0003$  & 0.0149 & $-0.7815$ && \\
$\langle^1P^\prime_1||t^1||^3P_0\rangle$ & $-0.5276$ & $-0.1885$ & 0.0009 & $-0.0006$  & 0.0003  & $-0.0185$ & $-0.2067$ && \\
$\langle^1S_0||t^1||^3S_1\rangle$ &  2.5679  & 2.1257 & 0.0 & 0.0024 & 0.0024  & 0.0042 & 2.1323 &&  \\
$\langle^1S_0||t^1||^3D_1\rangle$ &  0.0002  & $-0.1567$ & $-0.0004$ & 0.0 & $-0.0004$   & 0.0012 & $-0.1559$ &&  \\ \\
\hline
\multicolumn{10}{c}{M1 Reduced Matrix Elements} \\
$\langle^3P_1||m1||^3P_0\rangle$ & $-1.4131$  & $-1.2127$ & 0.0001    & $-0.0006$  & $-0.0005$  & 0.0 & $-1.2132$ & 1.413$^{s}$  & \\
$\langle^1P_1||m1||^3P_0\rangle$ &  0.0382    &   0.2684  & $-0.0002$ & $-0.0005$  & $-0.0007$  & 0.0 &   0.2677  & 0.042$^{s}$  & \\
$\langle^3P\prime_1||m1||^3P_0\rangle$ & $-0.1929$  &   0.0270    & $-0.0002$ & $-0.0008$  & $-0.0010$  & 0.0 & 0.0260 && \\
$\langle^1P^\prime_1||m1||^3P_0\rangle$& $-0.0451$  & $-0.2524$   & $-0.0001$ &   0.0004   & $-0.0003$  & 0.0 & $-0.2527$  && \\
$\langle^1S_0||m1||^3S_1\rangle$ & 0.0 & $-0.4584$  &   0.0005    &   0.0060  &   0.0065   &   0.0      & $-0.4519$ &&  \\
$\langle^1S_0||m1||^3D_1\rangle$ & 0.0 & $-0.0002$  &   0.0       &   0.0     &   0.0      &   0.0      & -0.0002 &&  \\
\end{tabular}
\end{ruledtabular}
\begin{flushleft}
  $^{\rm a}$ Ref. \cite{porsev-08}- CI+MBPT ,
  $^{\rm b}$ Ref. \cite{safronova-13}- CI+all-order,
  $^{\rm c}$ Ref. \cite{Husain-84},
  $^{\rm d}$ Ref. \cite{drozdowski-97},
  $^{\rm e}$ Ref. \cite{guo-10}- CI+CPMP,
  $^{\rm f}$ Ref. \cite{cooper-18}- CI+all-order,
  $^{\rm g}$ Ref. \cite{kelly-88},
  $^{\rm h}$ Ref. \cite{yasuda-06},
  $^{\rm i}$ Ref. \cite{glowacki-03} - CIDF,
  $^{\rm j}$ Ref. \cite{porsev-01} - MBPT,
  $^{\rm k}$ Ref. \cite{parkinson-76}
  $^{\rm l}$ Ref. \cite{wu-19}- DFCP+RCI,
  $^{\rm m}$ Ref. \cite{nagal-05},
  $^{\rm n}$ Ref. \cite{cheng-13} - CICP,
  $^{\rm o}$ Ref. \cite{vaeck-91} - MCHF,
  $^{\rm p}$ Ref. \cite{porsev-02} - MBPT,
  $^{\rm q}$ Ref. \cite{mitroy-10}- CI+MBPT,
  $^{\rm r}$ Ref. \cite{lu-23}- MCDHF,
  $^{\rm s}$ Ref. \cite{liu-07}- MCDF,
  $^{\rm t}$ Ref. \cite{kelly-80},
  $^{\rm u}$ Ref. \cite{jonsson-84},
  $^{\rm v}$ Ref. \cite{miller-92},
  $^{\rm w}$ Ref. \cite{redondo-04},
\end{flushleft}
\label{tab-e1m1}
\end{table*}

\begin{table}
\caption{E1, HFS and M1 reduced matrix elements (in a.u.) with increasing
	energy configurations in the model space. HFS reduced matrix elements
	are in terms of a factor of $10^{-7}$.}
\begin{ruledtabular}
\begin{tabular}{ccccc}
\multicolumn{1}{c}{\textrm{Transition}}  &
\multicolumn{1}{c}{\textrm{CF1$^{a}$}}  &
\multicolumn{1}{c}{\text{CF2$^{b}$}}&
\multicolumn{1}{c}{\text{CF3$^{c}$}}&
\multicolumn{1}{c}{\text{CF4$^{d}$}} \\
\hline
$\langle^1S_0||{\mathbf d}||^3P_1\rangle$ & $0.1033$  & $0.1169$  & $0.1059$ & 0.1404  \\
$\langle^1S_0||{\mathbf d}||^1P_1\rangle$ & $4.9905$  & $4.9892$  &  4.5568  & 4.8934  \\ \\

$\langle^3P_1||t^1||^3P_0\rangle$ &  2.4411  & 2.5027 & 2.5056 & 2.8449   \\
$\langle^1P_1||t^1||^3P_0\rangle$ &  2.0129  & 2.0367 & 2.0328 & 2.0534 \\ \\

$\langle^3P_1||m1||^3P_0\rangle$ & $-1.2308$  & $-1.2321$ & $-1.2134$ & $-1.2127$  \\
$\langle^1P_1||m1||^3P_0\rangle$ &  0.2897    &  0.2909   & 0.2889    & 0.2684   \\
\end{tabular}
\end{ruledtabular}
\begin{flushleft}
$^{\rm a}$ CF1: $5s^{2} + 5s5p$ \\
$^{\rm b}$ CF2: $5s^{2} + 5s5p + 5s4d$ \\
$^{\rm c}$ CF3: $5s^{2} + 5s5p + 5s4d + 5s6s$ \\
$^{\rm d}$ CF4: $5s^{2} + 5s5p + 5s4d + 5s6s + 5s6p$
\end{flushleft}
\label{tab-e1m1_2}
\end{table}

As evident from the table, and as can be expected, DF has the dominant 
contributions for all the matrix elements. To give an example from each category 
of reduced matrix elements, it contributes $\approx$ 95\%, 97\%, and 
117\% of the total value for $\langle^1S_0||{\mathbf d}||^1P_1\rangle$, 
$\langle^3P_1||t^1||^3P_0\rangle$  and $\langle^3P_1||m1||^3P_0\rangle$
matrix elements. The next dominant contributions come from the electron 
correlations due to residual Coulomb interaction. It contributes 
$\approx$ 4.6\%, 2.0\%, and $-16.5$\%, respectively, to these reduced 
matrix elements. As discernible from 
Fig. \ref{tbq-diag}, Breit, QED and perturbative triples too have the 
significant contributions. The largest contribution from the Breit is observed 
to be $\approx$ 1.0\% in the case of 
$\langle^1S_0||{\mathbf d}||^3P^\prime_1\rangle$, whereas a maximum QED 
correction of $\approx$ $-3.1$\% is observed in the case of 
$\langle^3P^\prime_1||m1||^3P_0\rangle$ reduced matrix elements. 
Perturbative triples are observed to have a large contribution of 
$\approx$ 8.9\% in the case of $\langle^1P^\prime_1||t^1||^3P_0\rangle$.

As can be observed from the table, there are several other theoretical 
and experimental studies on E1 reduced matrix elements in literature. 
The other theoretical values reported are using the methods like CI, MBPT, CI+MBPT, 
CI+all-order and different variations of MCDF. Among all these methods, 
in terms of accurate treatment of electron correlations, CI+all-order is closest to 
ours. There is, however, a key difference that it accounts for {\em core-core} 
and {\em core-valence} correlations only to the level of linear terms in CC.  
In FSRCC, however, we consider the nonlinear terms as well, which is crucial 
for multireference systems. For a comparative analysis, we consider the 
reduced matrix element $\langle^1S_0||{\mathbf d}||^1P_1\rangle$, which is 
expected to have a dominant contribution to the lifetime of the clock state. 
We notice a significant variation in the values reported from previous 
theory calculations and experiments for this. For example, the smallest value obtained 
using CI+CMP \cite{guo-10} defer by $\approx$ 10\% with the largest reported 
value using MCDF calculation \cite{vaeck-91}. Similarly, from 
experiments, there is a difference of $\approx$ 7\% between 
smallest \cite{yasuda-06} and largest \cite{kelly-80} reported values. 
Our calculated value, 4.88, is smallest among all the calculations, however, 
with more closer to the CI+all-order value \cite{safronova-13, cooper-18}. 
However, as evident from Table \ref{tab-e1m1_2} where we provide 
configuration-wise matrix elements, the reason for a smaller value is 
attributed to the more accurate treatment of {\em valence-valence} 
electron correlation, with the inclusion of higher energy configurations 
in the model space.
Our value, 4.99, with bare configurations $5s^2$ and $5s5p$ compares 
well with calculations \cite{safronova-13, cooper-18}, and in general 
with others, with a small difference due to relativistic, QED and 
perturbative triples corrections in our result. 
A similar trend of comparison and analysis also apply to all 
other E1 reduced matrix elements where our results are within the 
range of previous calculations and experiments, except 
for $\langle^1S_0||{\mathbf d}||^1P^\prime_1\rangle$ where 
our E1 amplitude is an order more than previous calculations.

To compare our results for HFS reduced matrix elements, to the best of our 
knowledge, we could find only one previous calculation \cite{lu-23}. 
Ref. \cite{lu-23} reports HFS reduced matrix elements for 
$^3P_1 \rightarrow {^3P_0}$ and $^1P_1 \rightarrow {^3P_0}$ transitions using 
MCDHF calculation. 
Our computed reduced matrix elements are larger by 
$\approx$ 42 and 31\%, respectively, with respect to Ref. \cite{lu-23}.
The reason for this is attributed to an accurate treatment of electron 
correlation effects in FSRCC theory. It should be noted that, MCDF 
suffers from an inherent dependency on the choice of configurations 
to incorporate the electron correlation effects, and therefore, does 
not provide an accurate treatment of {\em core-core} and 
{\em core-valence} correlations. 
The cumulative contributions from Breit and QED corrections from 
Ref. \cite{lu-23} are 0.35 and 0.41\%, respectively to the total value 
for the two transitions. We observed a slightly smaller contribution 
of $\approx$ 0.2\% for each reduced matrix elements.

For M1 reduced matrix elements, to the best of our knowledge, there is 
only one previous calculation \cite{liu-07}. Ref. \cite{liu-07} reports
M1 transition rates for $^3P_1\rightarrow{^3P_0}$ and $^1P_1\rightarrow{^3P_0}$ 
transitions using MCDF calculation. Our computed M1 reduced matrix element
for $^3P_1\rightarrow{^3P_0}$ is $\approx$ 17\% smaller than 
Ref. \cite{liu-07}. Similarly, we observed a very large difference of 
$\approx$ 85\% for $^1P_1\rightarrow{^3P_0}$ transition. Again, the reason 
for these differences could be attributed to inherent limitations with MCDF 
methods for accounting the electron correlations accurately in 
many-body calculations.

\subsection{Lifetime of Clock State}

Next, we present and discuss our results on the lifetime ($\tau$) of 
the clock state for Fermionic and Bosonic Sr using FSRCC theory. 
The $\tau$ from our calculations, along with other theory and 
experimental results, are provided in Table \ref{tab-tau}.  
As evident from the table, to assess the impact of {\em valence-valence} 
electron correlation, we have provided the configuration-wise 
contributions to lifetime. Moreover, for the same purpose, contributions
from the Breit interaction, QED effects and perturbative triples are 
also listed separately. 

\begin{table}
        \caption{Lifetime of the clock state, $5s5p{\;^3}P^{o}_0$, for ${^{87}}$Sr
        and ${^{88}}$Sr using FSRCC theory. Contributions from the Breit 
	interaction, QED effects, and perturbative triples are provided 
	separately for the quantitative assessment of the electron 
	correlations.}
\begin{center}
\begin{ruledtabular}
\begin{tabular}{lcr}
Conf./Methods  & \multicolumn{2}{c}{\textrm{$\tau$ (s)}} \\
           & ${^{87}}$Sr & ${^{88}}$Sr ($\times 10^{9}$)  \\
        \hline
MS1                &  $292.99$ & $2.918$ \\
MS2                &  $162.99$ & $2.568$ \\
MS3                &  $152.29$ & $2.918$ \\
MS4                &  $149.74$ & $9.283$ \\
Total CCSD         &  $149.74$ & $9.283$   \\
CCSD(T)            &  $143.23$    & $10.200$  \\
CCSD(T)+Breit      &  $144.58$    & $9.760$   \\
CCSD(T)+Breit+QED  &  $144.25$    & $11.046$  \\
Reco.              &  $144.25$    & $11.046$ \\
Oth. th.           &  110(30)$^{a}$, 156(9)$^{b}$, & 181.82$^{a}$  \\
                   &  132$^{c}$, 145(40)$^{d}$   &          \\
Expt.              &  118(3)$^{e}$, 330(140)$^{f}$,  &   \\
		   &  167$^{+79}_{-40}$$^{g}$, 151.4(48)$^{h}$  &   \\
\end{tabular}
\end{ruledtabular}
\begin{flushleft}
$^{\rm a}$ Ref. \cite{santra-04}- Model potential,
$^{\rm b}$ Ref. \cite{lu-23}- MCDF,
$^{\rm c}$ Ref. \cite{porsev-04}- CI + MBPT,
$^{\rm d}$ Ref. \cite{boyd-07}- Modified Breit-Wills theory,
$^{\rm e}$ Ref. \cite{muniz-21}- Expt.,
$^{\rm f}$ Ref. \cite{dorscher-18}- Expt.,
$^{\rm g}$ Ref. \cite{dolde-25}- Expt.,
$^{\rm h}$ Ref. \cite{lu-24}- Expt.
\end{flushleft}
\end{center}
\label{tab-tau}
\end{table}

\begin{figure}
\includegraphics[scale=0.30, angle=-90]{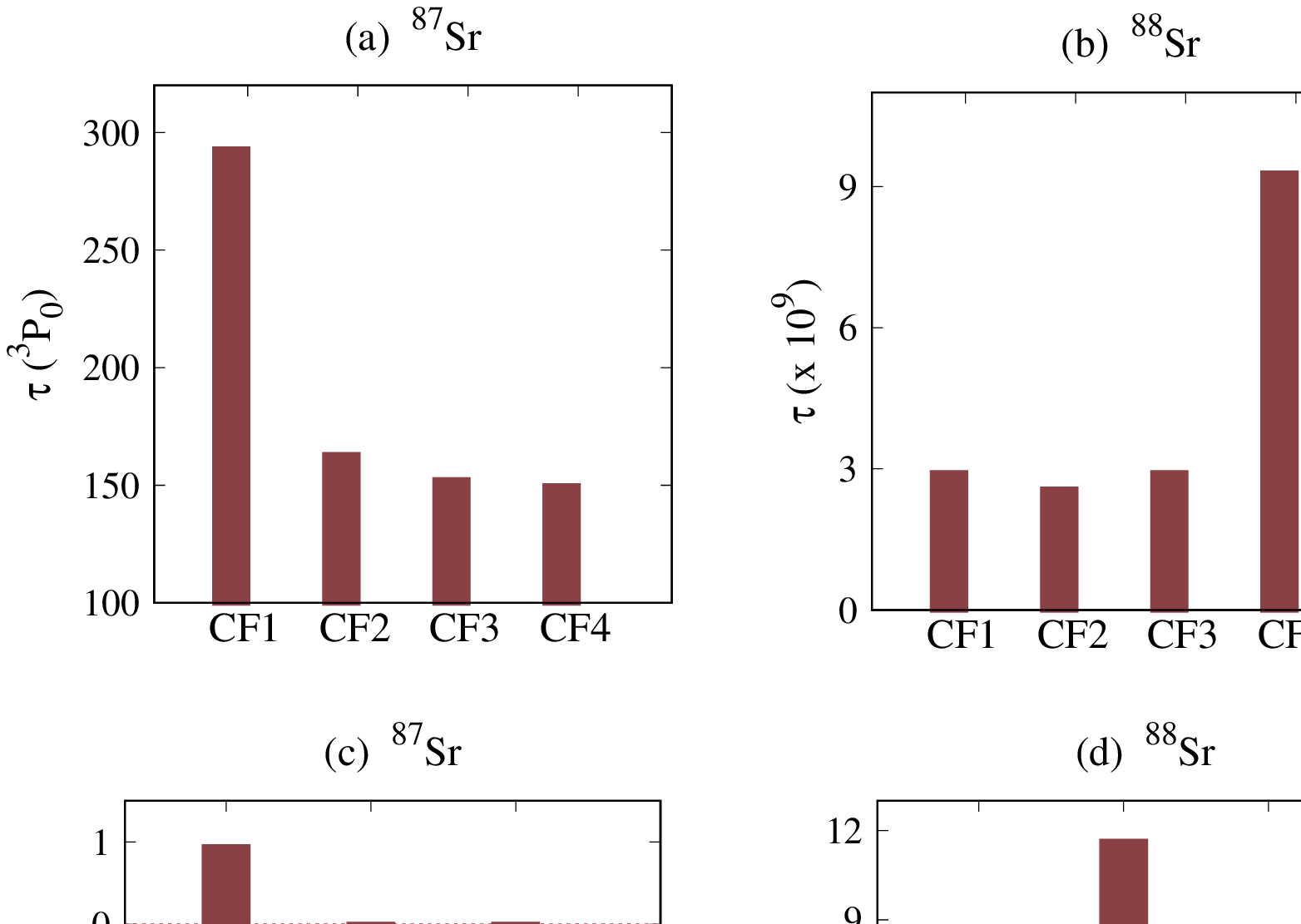}
	\caption{Percentage cumulative contributions to 
	the lifetime of the clock state from (a), (b) different high energy 
configurations in model space 
for ${^{87}}$Sr and ${^{88}}$Sr, respectively.  
(c), (d) Breit interaction, QED effects and perturbative triples to the lifetime 
of clock state for ${^{87}}$Sr and ${^{88}}$Sr, respectively.}
\label{tau-diag}
\end{figure}

\subsubsection {$^{87}${\textbf{Sr}}}

Lifetime of the clock state of ${^{87}}$Sr is calculated as the inverse of 
the HFS induced transition rate expressed in Eq.(\ref{hfs-gamma}).
For this, we used the E1 and HFS reduced matrix elements from our calculations 
listed in Table \ref{tab-e1m1} and experimental energies from the NIST database. 
Fig \ref{tau-diag} shows the cumulative percentage contributions from increasing 
model space, Breit and QED effects, and perturbative triples to the lifetime. 
As discernible from the panel (a) of the figure, there is a significant change 
in $\tau$ with the inclusion of high energy configurations. We observed a large 
change of $\approx$ 49\% when the model space is augmented from bare 
configuration MS1 to an extended configuration MS4. This highlights the 
importance of {\em valence-valence} electron correlations in multireference 
systems. The other significant contribution is from the perturbative triples. 
We observed a contribution of more than 4\% to $\tau$ with respect 
to CCSD value (panel (c)). The cumulative contribution from Breit and 
QED is observed to be $\approx$ 1\% of the total $\tau$ (panel (c)).

To compare our result with other theory calculations, we could find four results 
on $\tau$ of ${^{87}}$Sr. None of these are, however, using accurate methods 
like RCC. The other key point to observe is that, there is a significant variation 
in the $\tau$ reported in these calculations. For example, the smallest 
reported $\tau$ using model potential \cite{santra-04} is smaller 
by $\approx$30\% than the largest reported value using MCDF calculation \cite{lu-23}. 
The reported $\tau$ using CI+MBPT \cite{porsev-04} and Modified Breit-Wills theory \cite{boyd-07} 
are in between these two calculations. Moreover, as can be expected due to 
the nature of many-body methods employed, there are large errors in these 
calculations. A similar trend of significant variation and large 
errors is also observed in the experimental values. The reported values 
are in the range of 118 to 330 s, with a largest error of 47\% in recent 
experiment \cite{dolde-25}.
Our calculated $\tau$ is within the error bars of  previous theory 
and experimental values, and more closer to the CI+MBPT and 
Modified Breit-Wills theory calculations.

\subsubsection{$^{88}Sr$}

The lifetime of the clock state of ${^{88}}$Sr is calculated using 
E1 and M1 reduced matrix elements from our calculations, provided 
in Table \ref{tab-e1m1}, and excitation energies from NIST database 
in Eq. (\ref{e1m1-gamma}). As discernible from the panel (b) of 
the Fig. \ref{tau-diag}, unlike the case of ${^{87}}$Sr, here we observed 
a different trend of {\em valence-valence} electron correlation where 
MS4 is observed to have the largest change of $\approx$ 218\% with 
respect to MS3. The reason for this could be attributed to a strong 
mixing of $5s6p$ valence configuration with bare configuration $5s5p$. 
Consistent with ${^{87}}$Sr, but a larger contribution of $\approx$ 10\% of 
the CCSD value was observed from the perturbative triples.
Similarly, Breit and QED effects are observed to contribute 
more. The Breit contribution is $\approx$ -4\%, and reduces 
the total $\tau$, whereas QED effects increase $\tau$ by $\approx$ 12\%. 
It should be emphasized that these are significant contributions 
and can not be ignored. To compare our result with literature, 
to the best of our knowledge, we could find only one theoretical 
calculation for the lifetime, using the model potential calculation \cite{santra-04}. 
Our FSRCC value, $11.05\times10^{9}$ s, is smaller by an order 
of magnitude. 
The reason for this discrepancy could be attributed to the fact 
that Ref. \cite{santra-04} is based on a model potential calculation, 
whereas the present work uses a fully {\em ab initio} calculation.

\subsection{Dipole Polarizability}

The electric dipole polarizability of an atom or ion is a crucial parameter 
which governs the interaction with external electric fields, and therefore, can 
provide insights into various fundamental as well as technological 
implications \cite{khriplovich-91, griffith-09, karshenboim-10}.
In the context of atomic clocks, $\alpha$ is used in estimating the 
shift in the clock transition frequency due to black-body radiation (BBR).
The BBR shift in clock transition frequency is one of the most 
dominant frequency shifts for clocks at room temperature, and therefore, 
contribute heavily to the errors in the clocks. 
The $\alpha$ for an atom or ion can be computed using the expectation 
value of the dipole operator as
\begin{equation}
  \alpha = -\frac{\langle \widetilde \Psi_0|{\mathbf D}|\widetilde
         \Psi_0\rangle} {\langle \widetilde \Psi_0|\widetilde \Psi_0\rangle},
\label{alpha}
\end{equation}
where $|\widetilde \Psi_0\rangle$ is ground state perturbed wavefunction of Sr, 
obtained using the PRCC theory \cite{chattopadhyay-12, chattopadhyay-14, 
chattopadhyay-15, ravi-20, ravi-21b}.

\begin{figure}
\includegraphics[scale=0.30, angle=-90]{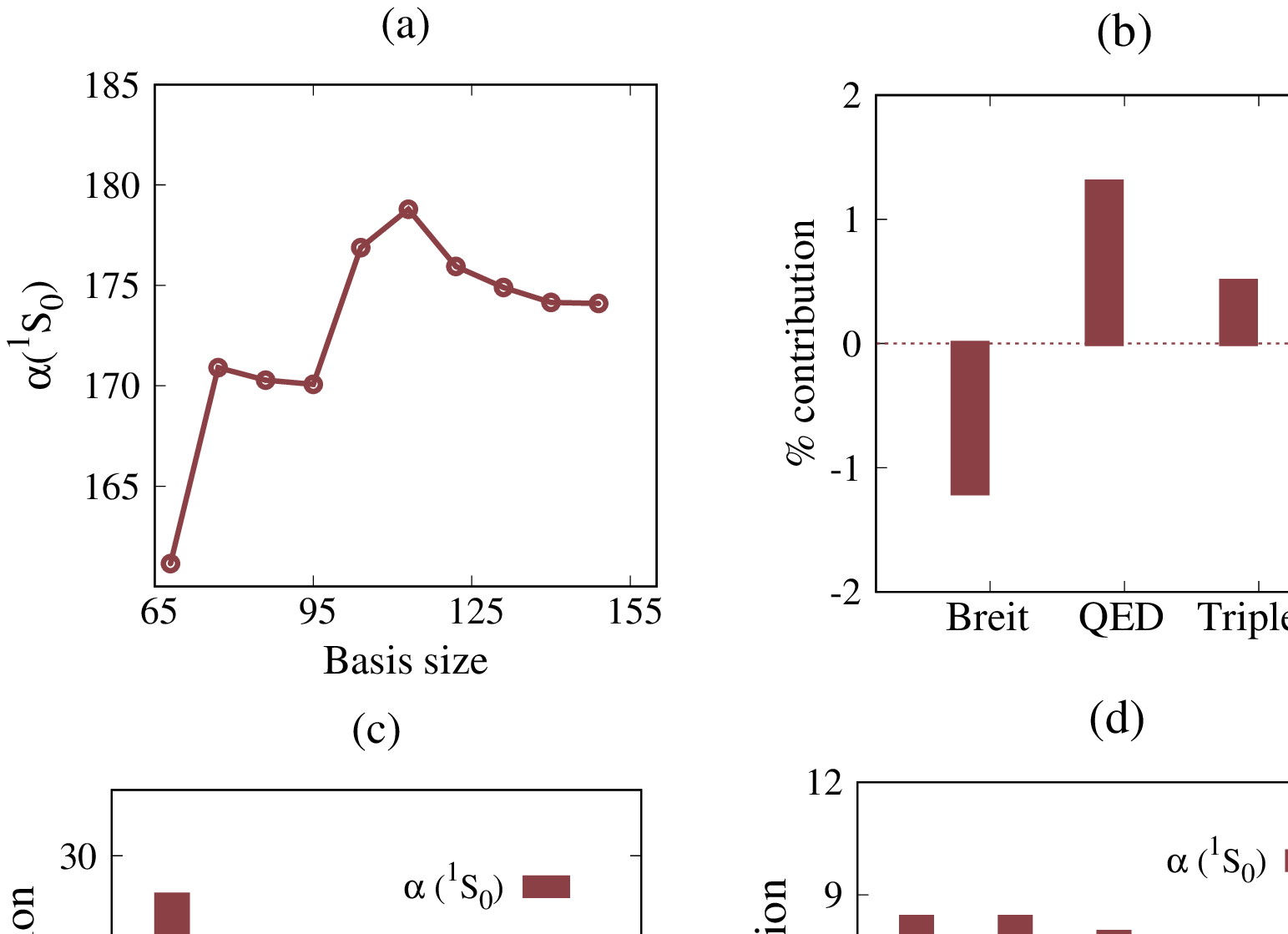}
\caption{(a) The convergence of $\alpha$ as a function of basis size, (b)
contributions from the various corrections to the polarizability, (c)
five largest percentage contributions from core orbitals to LO term
T$_1^{(1)\dagger}$D,  and (d) five leading contributions to NLO term
T$_{2}^{(1)\dagger}$DT$_{2}^{(0)}$ (in a.u.) from virtual-virtual orbital pairs.}  
\label{pol-diag}
\end{figure}

In Table \ref{tab-alpha}, we present our computed $\alpha$ for 
the ground state of Sr using PRCC theory. For the analysis of electron 
correlations, we have provided separate contributions at the different 
levels of the theory.
The contributions from the Breit, QED, and perturbative triples 
are also listed separately. The term {\em estimated} represents the 
cumulative contribution from the orbitals from $h$, $i$ and $j$-symmetries.
As evident from the table, and as can be expected, the largest contribution 
to $\alpha$ is from the DF.  It contributes about 107\% of the recommended value. 
The DF contribution is computed by replacing PRCC wavefunction 
$|\widetilde \Psi_0\rangle$ in Eq. (\ref{alpha}) with a DF wavefunction. 
The linearized PRCC is observed to increase $\alpha$ by 136\% of DF value
due to electron correlation effects. The nonlinear terms are observed to 
have a significant opposite contribution of $\approx$ 18\%, reducing 
the $\alpha$ below LPRCC. As discernible from Fig. \ref{pol-diag}(a), 
$\alpha$ is well converged with respect to basis size. 
The Breit, QED and perturbative triples are observed to have important 
contributions of $-1.2$\%, $1.3$\% and $0.5$\%, respectively, of the total value. 
Our recommended value, 176.5 a.u, of $\alpha$ is within the experimental 
error bar of the value 186$\pm$18 a.u. reported in Ref. \cite{schwartz-74}.

\begin{table}
\caption{The value of $\alpha$ (a.u.) for ground state, $5s^2\;^1S_{0}$,
of Sr using PRCC theory. The data available from experiment and other 
theory calculations are also provided for the comparison.}
\begin{ruledtabular}
\begin{tabular}{cc}
Method   & $\alpha$   \\
\hline
DF                &  $156.826$ \\
LPRCC             &  $213.429$    \\
PRCC              &  $174.893$ \\
PRCC(T)           &  $175.773$ \\
PRCC(T)+Breit     &  $173.674$ \\
PRCC(T)+Breit+QED &  $175.597$ \\
Estimated         &  $176.471$ \\
Recommended       &  $176.50 \pm 2.65$ \\
Other cal.        & $197.8^{\rm a}$, $198.85^{\rm b}$, $202^{\rm c}$, $197.14(20)^{\rm d}$,    \\
                  & $193(13)^{\rm e}$ , $190^{\rm f}$, $197.2^{\rm g}$, $201.2^{\rm h}$,        \\
                  & $198.9^{\rm i}$,  $193.2^{\rm j}$, $190.82^{\rm k}$,  \\
                  & $202.02^{\rm l}$, $186.98(85)^{\rm m}$ \\
Expt.             & $186(15)^{\rm n}$ \\
\end{tabular}
\end{ruledtabular}
\begin{tabbing}
  $^{\rm a}$Ref.\cite{cheng-13}-CICP,  \\
  $^{\rm b}$Ref.\cite{lim-99}- RCCSD,   \\
  $^{\rm c}$Ref.\cite{porsev-06}-CI+MBPT, \\
  $^{\rm d}$Ref.\cite{safronova-13}-  CI+all-order, \\
  $^{\rm e}$Ref.\cite{guo-10}- B-spline configuration interaction with a  semi-empirical \\ core-polarization model potential, \\
  $^{\rm f}$Ref.\cite{sadlej-91}-CCSD, \\
  $^{\rm g}$Ref.\cite{porsev-08}-CI+MBPT, \\
  $^{\rm h}$Ref.\cite{mitroy-03}- Semi-empirical approach using model potential, \\
  $^{\rm i}$Ref.\cite{lim-04}-DK-CCSD(T), \\
  $^{\rm j}$Ref.\cite{patil-00}-Model potential with
  a hard core and the correct large-r \\ coulombic behavior, \\
  $^{\rm k}$Ref.\cite{chattopadhyay-14}-PRCC, \\
  $^{\rm l}$Ref.\cite{wu-19}-DFCP+RCI, \\
  $^{\rm m}$Ref.\cite{singh-13}-CCSD \\
  $^{\rm n}$Ref.\cite{schwartz-74}-Expt., \\
\end{tabbing}
\label{tab-alpha}
\end{table}

In Table \ref{tab-alpha} we compare $\alpha$ from previous theory calculations 
and experiment. The ground state 
$\alpha$ for Sr has been calculated using CI \cite{wu-19}, CICP \cite{cheng-13}, 
CI+MBPT \cite{porsev-06, porsev-08}, CI+all-order \cite{safronova-13}, 
and the coupled-cluster \cite{lim-99,sadlej-91,lim-04, sahoo-08, chattopadhyay-14, singh-13} 
method such as ours.
However, as evident from the table, there is a variation in the results using 
different methods. The average value of $\alpha$ reported in previous 
calculations other than Ref. \cite{singh-13} is 197 a.u., which is higher than 
the experimental value \cite{schwartz-74} by $\approx$ 5\%. The $\alpha$, 186.98, reported 
in Ref. \cite{singh-13} is closest to the experiment. In terms of methodology, among 
all the previous calculations, our work is more closer to Ref. \cite{singh-13}, 
however, with a key difference in the implementation of Breit and QED corrections. 
Our recommended value, 176.5 a.u., is $\approx$ 5\% 
smaller than Ref. \cite{singh-13}.  Our DF value, 156.827 a.u., is however in 
excellent agreement with the value, 156.83 a.u., reported in Ref. \cite{singh-13}. 
The Breit and QED corrections from our work are observed to be larger than 
the corrections $-0.09$ and $-0.01$\%, respectively, reported in Ref. \cite{singh-13}.
In addition, the inclusion of perturbative triples in our work accounts 
for $\approx$ 0.5\% of the total value. The reason for the difference 
in $\alpha$ from our previous calculation \cite{chattopadhyay-14} is 
attributed to the inclusion of nonlinear terms in PRCC theory in the present work.
The difference in the $\alpha$ value from CI+MBPT results \cite{porsev-06, porsev-08} 
could be attributed to the treatment of {\em core-core} electron correlation 
to all orders of residual Coulomb interaction in the present work. 
The CI + all-order calculation \cite{safronova-13} accounts for these 
correlations to all orders, however, using a linearized 
CC theory, whereas the present work uses a nonlinear CC theory to account for 
the effects of residual Coulomb interaction. This explains the reason for the difference 
in $\alpha$ values from present work and CI + all-order calculation \cite{safronova-13}.

In Table \ref{pol_tw} we provide the contributions from various correlation 
terms in PRCC theory. The leading order (LO) contribution is observed from the 
term ${\mathbf T}_1^{(1)\dagger}D$ + H.c., where it contributes $\approx$ 115\% 
of the final value. The large contribution is expected, as this term includes 
the contributions from DF and the dominant core-polarization effects. As discernible from 
Fig. \ref{pol-diag}(c), $\approx$ 48\% of the LO contribution comes 
from the $5s$ valence electrons via the dipolar mixing with $6p$ states. 
In the remaining, $\approx$ 34\% and 9\% of contributions arise from 
the dipolar mixing with $5p$ and $7p$ states, respectively. 
The next-to-leading order (NLO) term, T$_{2}^{(1)\dagger}$DT$_{2}^{(0)}$, 
contributes $\approx$ 11\% of the total value, highlighting the importance
of pair-correlation effects in closed-shell systems.
As discernible from Fig. \ref{pol-diag}(d), we observed dominant contributions 
from $5p - 6p$ and $6p - 7p$ virtual-virtual pairs.

\begin{table}
    \caption{Termwise contributions to $\alpha$ (a.u.) from different 
	terms in PRCC theory.}
    \begin{center}
    \begin{ruledtabular}
    \begin{tabular}{cc}
        Terms + h. c. & $\alpha$  \\
        \hline                                                 
        ${{\mathbf T}_{1}^{(1)\dagger}{\mathbf D}} $              &  $201.4959$  \\
        ${\mathbf T_{1}}{^{(1)\dagger}}{\mathbf D}T_{2}^{(0)} $   &  $-11.9852$ \\
        ${\mathbf T_{1}}{^{(1)\dagger}}{\mathbf D}T_{1}^{(0)} $   &  $-17.2535$   \\
        ${\mathbf T_{2}}{^{(1)\dagger}}{\mathbf D}T_{1}^{(0)} $   &  $10.0744$    \\
        ${\mathbf T_{2}}{^{(1)\dagger}}{\mathbf D}T_{2}^{(0)} $   &  $18.5295$     \\
        Normalization                             &  $1.1485$      \\
        Total                                     &  $174.8931$       \\
    \end{tabular}
    \end{ruledtabular}
    \end{center}
    \label{pol_tw}
\end{table}

\subsection{Isotope shift}

{\bf 
\begin{table*}[t]
\small 
\centering
\caption{The normal mass shift (GHz-amu), specific mass shift (GHz-amu) and field
	shift (MHz fm$^{-2}$) factors of Sr calculated using MCDF and FSRCC theories. 
	We have also provided the data from other theory calculations and experiments 
	for comparison.} 
\label{iso_tab}
\begin{ruledtabular}
\begin{tabular}{cccccclccl}
State/Transition  & \multicolumn{3}{c}{$K^{\rm NMS}$} &
\multicolumn{3}{c}{$K^{\rm SMS}$} & 
\multicolumn{3}{c}{$F_{\rm s}$}  \\ 
\cline{2-4} \cline{5-7} \cline{8-10}
& MCDF & FSRCC & Oth. results & MCDF & FSRCC & Oth. results & MCDF & FSRCC & Oth. results \\
\hline 
$^1S_0$ - $^3P_0$ & $-237.02$ & $-405.72$  & $-235.33^{a}$ & $-323.61$ & $-576.67$ & & $-812.01$ & $-1260.43$ & \\
                 \\
$^1S_0$ - $^3P_1$ & $-239.94$ & $-352.06$  & $-238.63^{a}$ & $-322.92$ & $-20.03$  & $191.1^{b}$,
	          & $-814.34$ & $-1240.95$ & $-1034 \pm 24^{b}$,         \\
                  &  &  &     & & & $189.96 \pm 12.87^{c}$,  &  &   & $-953.18 \pm 67.42^{c}$,  \\
                  &  &  &     & & & $-469.22^{d}$,           &  &   & $-1055.36^{d}$,            \\
                  &  &  &     & & & $-0.151^{e}$,            &  &   & $-955.06^{e}$,              \\
                  &  &  &     & & & $190.90 \pm 119.32^{f}$  &  &   & $-1106 \pm 165^{f}$          \\
                          \\

$^1S_0$ - $^1P_1$ & $-357.79$ & $-332.62$  & $-357.12^{a}$,& $-104.37$ & $47.39$   & $-41.62 \pm 20.81^{c}$, & $-764.35$ & $-1234.55$ & $-788.39 \pm 102.99^{c}$, \\
                  &  &   & $-357.21^{g}$,&  &  & $-211.90^{e}$,           &   & & $-936.33^{e}$,             \\
                  &  &   & $356.9^{i}$   &  &  & $-42.85 \pm 28.57^{f}$,  &   & & $-896 \pm 130^{f}$ ,        \\
                  &  &   &               &  &  & $68.112^{g}$,            &   & & $-806 \pm 36^{g}$,           \\
                  &  &   &               &  &  & $-67.85 \pm 53.57^{h}$,  &   & & $1579 \pm 47^{h}$,            \\
                  &  &   &               &  &  & $-23 \pm 3^{i}$          &   & & $-628 \pm 64^{i}$              \\
\end{tabular}
\end{ruledtabular}
\begin{tabbing}
	$^{\rm a}$Semi-empirically derived result using the relation $K^{\rm NMS}$ = $\nu$/1822.9, where $\nu$ is the experimental transition frequency,  \\
  $^{\rm b}$Ref.\cite{bushaw-97}- Expt.,  \\
  $^{\rm c}$Ref.\cite{bender-84}- Expt.,   \\
  $^{\rm d}$Ref.\cite{chidichimo-85}- First-order perturbative treatment, \\
  $^{\rm e}$Ref.\cite{aspect-91}- Multiconfiguration Hartree-Fock (MCHF) theory, \\
  $^{\rm f}$Ref.\cite{buchinger-85}- Expt., \\
  $^{\rm g}$Ref.\cite{foot-84}- Expt., \\
  $^{\rm h}$Ref.\cite{buchinger-87}- Expt., \\
  $^{\rm i}$Ref.\cite{bushaw-00}- Expt. \\
\end{tabbing}
\label{is_factors}
\end{table*}

}

In Table \ref{is_factors}, we present our FSRCC results on normal mass shift
($K^{\rm NMS}$), specific mass shift ($K^{\rm SMS}$), and field
shift ($F_{\rm s}$) factors for the clock transition ($5s^2\ ^1S_0
\rightarrow 5s5p\ ^3P_0$) and two intercombination lines
($5s^2\ ^1S_0 \rightarrow 5s5p\ ^3P_1$ and
$5s^2\ ^1S_0 \rightarrow 5s5p\ ^3P_1$) of Sr. In addition, awe also compute and
provide the results from MCDF calculations for analysis and comparison of
electron correlation effects.
Since MCDF and FSRCC treat electron correlation in fundamentally different ways,
their comparison provides valuable insights into the interplay of many-body
effects in Sr. The FSRCC theory, being an all-order relativistic many-body
approach, serves as a rigorous benchmark for assessing the accuracy and
reliability of the MCDF results.

The MCDF results were obtained using RIS4 \cite{ekman-19} module interfaced
with GRASP2K\cite{jonsson-13}.
We accounted for {\em core-valence} and {\em valence-valence} correlations
by allowing single and double replacements of the electrons.
While {\em valence-valence} correlation describe the coupling among outer shell
electrons, {\em core-valence} correlation captures the interplay between inner
and outer shells. To achieve a balanced description of correlation effects,
a multireference framework was adopted, where we used dominant configurations
\{5s$^{2}$, 5p$^{2}$, 5s4d\} (even-parity) and \{5s5p, 4d5p\} (odd-parity)
in the configuration space. The active orbital space was extended
up to \{12s12p7d\} for even- and \{12s12p6d\} for odd-parity configurations.
The FSRCC results were, however, obtained using an in-house code \cite{palki-25}.
For this, we used a converged basis set with 131 orbitals
($19s$, $17p$, $17d$, $14f$, and $8g$) in active space and a model space
restricted to $5s^2$ and $5s5p$ configurations to ensure the computational
tractability while retaining essential correlation effects.

For NMS, to the best of our knowledge, there are no experimental results for
$^1S_0$ - ${^3P_0}$ and $^1S_0$ - ${^3P_1}$ transitions. However, for $^1S_0$ - ${^1P_1}$
transition, we could find two experimental results \cite{foot-84, bushaw-00}.
As evident from the table, the NMS factor obtained from our MCDF calculations
show a good agreement with the semi-empirical value for all the transitions.
However, our FSRCC results are different than both MCDF and semi-empirical
values. The reason for this is attributed to the more accurate treatment of
electrons correlations in FSRCC theory.

For the FS factor, as can be observed from the table, there is a large variation
in the reported experimental data, and moreover, they have large errors.  We find,
the differences of $\approx$ 16\% and 151\% in the smallest and largest reported
experimental values for $^1S_0$ - ${^3P_1}$ and $^1S_0$ - ${^1P_1}$ transitions,
respectively. Our MCDF results for $^1S_0$ - ${^3P_1}$ and $^1S_0$ - ${^1P_1}$
transitions differ from that of Ref. \cite{aspect-91} by $\approx$ 15\%
and 18\%, respectively. The reason for this could be attributed to the differences
in the configuration space considered and correlation treatment. Our FSRCC result for
$^1S_0$ - ${^3P_1}$ transition lies within the experimental uncertainty of
reported value in Ref. \cite{buchinger-85}. For $^1S_0$ - ${^1P_1}$ transition,
like the case of $^1S_0$ - ${^3P_1}$, it is consistent with \cite{buchinger-85},
however, in general on the higher side of other experimental values.

The calculation of SMS factor is the most challenging among all the three
isotope shifts factors. The reason for this is the two-body nature of the
SMS operator, which complicates the angular momentum coupling for two-valence
systems. As evident from the table, the reported SMS factors from previous
calculations and experiments exhibit large variations in both sign and
magnitude. Also, the reported experimental values are from the old
measurements and have large errors.
For $^1S_0$ - ${^3P_1}$ transition, our MCDF value $-322.92$ GHz-amu have the same sign but
smaller in magnitude than the calculation \cite{chidichimo-85}. On contrary, among
all the reported results, our FSRCC result, $-20$ GHz-amu, is more closer to
the MCDF calculation \cite{aspect-91}
For $^1S_0$ - ${^1P_1}$, in terms of magnitude, the FSRCC result is more closer
to the experiments than the MCDF values, demonstrating the superior accuracy
of the FSRCC approach. Additionally, the sign of our SMS factor is consistent
with experiment \cite{foot-84}. The close agreement between the FSRCC results
and the best available experimental data highlights the robustness and
predictive capability of the FSRCC method for high-precision isotope
shift studies.

\section{Theoretical Uncertainty}

As given in Eqs. (\ref{e1hfs}) and (\ref{e1m1-gamma2}), theoretical uncertainty 
in the computed $\tau$ depends on the uncertainties in the E1 and M1 matrix 
elements, and the energy denominators. Since there are accurate data on the 
excitation energies from experiments, we have used them in the calculation 
of lifetime. As experimental results are not available for all the E1 and 
M1 reduced matrix elements, we have identified four different sources which 
can contribute to the uncertainties in these matrix elements. The first source 
of uncertainty is due to the truncation of the basis set in our 
calculation. As discussed in the basis convergence section, our calculated 
values of E1 and M1 reduced matrix elements converge well to the order 
of $10^{-3}$ or smaller with basis. Since this is a very small change, 
we can neglect this uncertainty. The second source of uncertainty arises 
due to the truncation of the dressed Hamiltonian at the second order of 
$T^{(0)}$ in the properties calculation \cite{mani-11, ravi-21b}. 
In our earlier work \cite{mani-10}, using an iterative scheme, we found 
that the terms with third and higher orders in $T^{(0)}$ contribute less 
than 0.1\%. So, we consider 0.1\% as an upper bound from this source 
of uncertainty. The third source is due to the partial inclusion of 
triple excitations in the FSRCC theory. Since the perturbative triples 
account for the leading order terms in triple excitation, the contribution 
from remaining terms will be small. 
Based on the analysis from our previous works \cite{ravi-20, chattopadhyay-15},
we estimate the upper bound from this source of uncertainty as 0.72\%. 
The fourth source of uncertainty could be associated with the frequency-dependent 
Breit interaction which is not included in the present calculation. However, 
in our previous work \cite{chattopadhyay-14}, using a series of computations 
using GRASP2K we estimated an upper bound on this uncertainty as 0.13\% in Ra. 
So, for the present work, we take 0.13\% as an upper bound from this source. 
There could be other sources of theoretical uncertainty, such as the 
higher order coupled perturbation of vacuum polarization and self-energy terms, 
quadruply excited cluster operators, etc. However, in general, these all 
have much lower contributions to the properties and their cumulative 
theoretical uncertainty could be below 0.1\%. 

The other source of theoretical uncertainty which will contribute to the 
lifetime of the clock state is the QED corrections at the level 
of Eqs. (\ref{e1hfs}, \ref{e1m1-gamma2}). To estimate this uncertainty, we refer to 
the Refs. \cite{shabaev-05a, shabaev-05b}.  In these works, Shabaev and 
collaborators have computed the one-loop QED corrections to the properties 
of Cs and Fr. The reported total contribution is about 0.28\%. So, based on 
these works we consider 0.3\% as the upper bound from this source of uncertainty. 
By combining the upper bounds of all the uncertainties, the theoretical 
uncertainty associated with the lifetime of the clock state is $\approx$ 3.5\%. 
It should, however, be noted that the uncertainty in the value of $\alpha$ is 
much smaller, about 1.5\% \cite{ravi-22}

\section{Conclusions}

We have employed an all-particle multireference FSRCC theory to investigate the
clock transition properties in both Fermionic and Bosonic isotopes of Sr.
We have computed the excitation energies, E1 and M1 matrix elements, and the
lifetime of the metastable clock states in $^{87}{\rm Sr}$ and
$^{88}{\rm Sr}$. To account for valence-valence electron correlation more
accurately using FSRCC, we used a larger model space consisting of
$5s^2$, $5s5p$, $5s4d$, $5s6s$, and $5s6p$ configurations.
Moreover, we have implemented the calculation of isotope shifts
within the framework of FSRCC theory and have computed isotope shift
parameters for clock transition and two intercombination lines in Sr.
Furthermore, we employed a PRCC theory to compute the ground state electric
dipole polarizability of Sr. To improve the accuracy of the computed properties,
we incorporated the corrections from the Breit interaction, self-energy correction,
vacuum polarization and perturbative triples to all our calculations. Furthermore,
the convergence of the properties were ensured using large bases.

Our calculated excitation energies of low-lying states show a good agreement
with experimental results. Our calculated E1, M1 and HFS matrix elements are
consistent with previous calculations, however, slightly different because
of more accurate treatment of electron correlations in our calculations.
Using our calculated E1, M1 and HFS reduced matrix elements, we calculated
the lifetime of the metastable clock states for $^{87}{\rm Sr}$ and
$^{88}{\rm Sr}$. Our calculated lifetime for $^{87}{\rm Sr}$ is within
the error bar of the experimental results. It should, however, be mentioned
that experiments and previous calculations have large errors. For $^{88}{\rm Sr}$,
our calculated lifetime is an order of magnitude smaller than the only
available calculation \cite{santra-04} using model potential.
Our recommended value for ground state $\alpha$ is within the experimental
error bar. As can be expected due to more accurate treatment of electron 
correlations, our FSRCC results on isotope shifts parameters show differences 
from the MCDF calculations.

From the detailed analysis of electron correlations, we find that the corrections
from the Breit interaction, QED effects, and perturbative triples are crucial
to get accurate lifetimes of the clock states. The cumulative contributions
from Breit interaction and QED effects to the lifetime are observed to be $\approx$ 9\% and 1\%
for $^{88}$Sr and $^{87}$Sr, respectively. Whereas, perturbative triples are
found to contribute $\approx$ 8.3\% and 4.5\% for $^{88}$Sr and
$^{87}$Sr, respectively. Our calculations also show that, perturbative triples
are crucial to get accurate excitation energies.

\begin{acknowledgments}

The authors wish to thank Ravi Kumar and Suraj Pandey for the useful discussion. Palki
acknowledges the fellowship support from UGC (BININ04154142), Govt. of India. B. K. M
acknowledges the funding support from SERB, DST (CRG/2022/003845). Results presented 
in the paper are based on the computations using the High Performance Computing 
clusters Padum at IIT Delhi and PARAM RUDRA facility at IUAC under the National 
Supercomputing Mission of Government of India.
\end{acknowledgments}

\appendix
\section{Convergence table of the properties with basis size}

In Table  \ref{conv_tab}, we have shown the convergence trend for E1, HFS
and M1 reduced matrix elements for all the transitions considered in the
present work. Similarly, Table \ref{conv-alpha} shows convergence of 
dipole polarizability with basis size.

\begin{sidewaystable*}[p]
\centering
\vspace*{9.5cm}
        \caption{Convergence trend of E1, HFS, and M1 reduced matrix elements with 
 	basis size.}
        \begin{ruledtabular}
        \begin{tabular}{ccccccccccccccccccc}
	       
                 \multicolumn{1}{c}{\textrm{States}}  &
                 \multicolumn{1}{c}{\text{BS-68$^{a}$}}&
                 \multicolumn{1}{c}{\text{BS-77$^{b}$}}&
                 \multicolumn{1}{c}{\text{BS-86$^{c}$}}&
                 \multicolumn{1}{c}{\text{BS-95$^{d}$}} &     
                 \multicolumn{1}{c}{\text{BS-104$^{e}$}} &  
                 \multicolumn{1}{c}{\text{BS-113$^{f}$}} &     
                 \multicolumn{1}{c}{\text{BS-122$^{g}$}} &
                 \multicolumn{1}{c}{\text{BS-131$^{h}$}} &   
                 \multicolumn{1}{c}{\text{BS-140$^{i}$}}  &
                 \multicolumn{1}{c}{\text{BS-149$^{j}$}}  &
                 \multicolumn{1}{c}{\text{BS-158$^{k}$}}  &
                 \multicolumn{1}{c}{\text{BS-167$^{l}$}}  &
                 \multicolumn{1}{c}{\text{BS-176$^{l}$}}  &
                 \multicolumn{1}{c}{\text{BS-185$^{l}$}}  &
                 \multicolumn{1}{c}{\text{BS-194$^{l}$}}  &
                 \multicolumn{1}{c}{\text{BS-203$^{l}$}}  \\

            \hline     \\ 

\hline \\
\multicolumn{16}{c}{E1 reduced matrix elements (in a.u.)} \\ \\

$\langle^1S_0|| \bm{d} ||^3P_1\rangle$    &  0.1495 &  0.1491 &  0.1484 &  0.1473 &  0.1472 &  0.1477 &  0.1482 &  0.1463 &  0.1445 &  0.1440 &  0.1436   
                                          &  0.1425 &  0.1409 &  0.1404 &  0.1402 &  0.1398         \\
$\langle^1S_0|| \bm{d} ||^1P_1\rangle$    &  5.0088 &  5.0056 &  5.0054 &  5.0064 &  4.9985 &  4.9809 &  4.9544 &  4.9396 &  4.9228 &  4.9109 &  4.8997       
                                          &  4.8965 &  4.8969 &  4.8934 &  4.8904 &  4.8894          \\
$\langle^1S_0|| \bm{d} ||^3P_1^{\prime}\rangle$ &  0.3848 &  0.3859 &  0.3860 &  0.3856 &  0.3829 &  0.3752 &  0.3597 &  0.3510 &  0.3453 &  0.3498 &  0.3471  
                                                &  0.3455 &  0.3422 &  0.3384 &  0.3369 &  0.3359            \\
$\langle^1S_0|| \bm{d} ||^1P_1^{\prime}\rangle$ & $-2.6055$ & $-2.6116$ & $-2.6137$ & $-2.6150$ & $-2.6072$ & $-2.6035$ & $-2.6117$ & $-2.6136$ & $-2.6136$ 
                                                & $-2.6139$ & $-2.6170$ & $-2.6189$ & $-2.6195$ & $-2.6213$ & $-2.6219$ & $-2.6222$    \\
$\langle^3S_1|| \bm{d} ||^3P_0\rangle$    &  2.0494 &  2.0447 &  2.0413 &  2.0388 &  2.0350 &  2.0265 &  2.0157 &  2.0037 &  1.9924 &  1.9835 &  1.9774 
                                          &  1.9748 &  1.9703 &  1.9686 &  1.9678 &  1.9673                \\
$\langle^3D_1|| \bm{d} ||^3P_0\rangle$    & $-2.5920$ & $-2.5884$ & $-2.5886$ & $-2.5898$ & $-2.5959$ & $-2.5822$ & $-2.5199$ & $-2.4742$ & $-2.4674$ & $-2.4658$ 
                                          & $-2.4559$ & $-2.4403$ & $-2.4347$ & $-2.4329$ & $-2.4321$ & $-2.4317$              \\ \\

\hline \\
\multicolumn{16}{c}{HFS Reduced Matrix Elements ($\times$ 10$^{-7}$)} \\ \\
$\langle^3P_1|| t^{1} ||^3P_0\rangle$    &  2.1052 &  2.1679 &  2.2239 &  2.2539 &  2.2841 &  2.3210 &  2.3773 &  2.4031 &  2.4215 &  2.4533 &  2.4789 &  2.4904  &  2.4849  &  2.4834  &  2.4789  &            \\          
$\langle^1P_1|| t^{1} ||^3P_0\rangle$    &  1.6078 &  1.6491 &  1.6833 &  1.7064 &  1.7263 &  1.7539 &  1.7941 &  1.8019 &  1.8211 &  1.8643 & 1.8830 & 1.8986  &  1.9099 &  1.8759 & 1.8277  &           \\ \\

\hline \\
\multicolumn{16}{c}{M1 Reduced Matrix Elements (in a.u.)} \\ \\
$\langle^3P_1|| m1 ||^3P_0\rangle$    & $-1.2035$  & $-1.2027$ & $-1.2022$ & $-1.2019$ & $-1.2054$ & $-1.2074$ & $-1.2083$ & $-1.2090$ & $-1.2112$ & $-1.2122$ & $-1.2129$
                                      & $-1.2125$  & $-1.2123$ & $-1.2127$ & $-1.2131$ & $-1.2132$     \\ 
$\langle^1P_1|| m1 ||^3P_0\rangle$    &  0.2829  &  0.2833 &  0.2835 &  0.2836 &  0.2806 &  0.2782 &  0.2759 &  0.2742 &  0.2716 &  0.2702 &  0.2692 &  0.2691 &  0.2688  
                                      &  0.2684 &  0.2679  &  0.2678      \\  
$\langle^3P_1^{\prime}|| m1 ||^3P_0\rangle$ &  0.0830    &  0.0818   &  0.0808   &  0.0800   &  0.0749   &  0.0679   &  0.0597   &  0.0523   &  0.0429   &  0.0370   &  0.0322 
                                            &  0.0312    &  0.0290   &  0.0270   &  0.0259   &  0.0253       \\ 
$\langle^1P_1^{\prime}|| m1 ||^3P_0\rangle$ & $-0.2609$  & $-0.2616$ & $-0.2619$ & $-0.2622$ & $-0.2598$ & $-0.2582$ & $-0.2576$ & $-0.2568$ & $-0.2548$ & $-0.2532$ & $-0.2525$
                                            & $-0.2526$  & $-0.2526$ & $-0.2524$ & $-0.2521$ & $-0.2519$         \\ 
$\langle^1S_0|| m1 ||^3S_1\rangle$          & $-1.0739$  & $-1.0730$ & $-1.0750$ & $-1.0772$ & $-0.9998$ & $-0.8399$ & $-0.6549$ & $-0.5614$ & $-0.5486$ & $-0.5324$ & $-0.5044$
                                            & $-0.4748$  & $-0.4616$ & $-0.4584$ & $-0.4573$ & $-0.4565$         \\ 
$\langle^1S_0|| m1 ||^3D_1\rangle$          & $-0.0002$  & $-0.0002$ & $-0.0003$ & $-0.0003$ & $-0.0003$ & $-0.0002$ & $-0.0002$ & $-0.0002$ & $-0.0002$ & $-0.0002$ & $-0.0002$ 
                                            & $-0.0002$  & $-0.0002$ & $-0.0002$ & $-0.0002$ & $-0.0002$          \\    

           \footnotetext{Yu. Ralchenko, A. Kramida, J. Reader, and the NIST ASD Team, NIST Atomic Spectra Database (version 4.1), available at http:physics.nist.gov/    asd (National Institute of Standards and Technology, Gaithersburg, MD, 2011)} 
           \footnotetext{Basis 68  - 12s, 10p, 10d, 7f, 1g}
           \footnotetext{Basis 77  - 13s, 11p, 11d, 8f, 2g}
           \footnotetext{Basis 86  - 14s, 12p, 12d, 9f, 3g}
           \footnotetext{Basis 95  - 15s, 13p, 13d, 10f, 4g}
           \footnotetext{Basis 104  - 16s, 14p, 14d, 11f, 5g}
           \footnotetext{Basis 113 - 17s, 15p, 15d, 12f, 6g}
           \footnotetext{Basis 122 - 18s, 16p, 16d, 13f, 7g}
           \footnotetext{Basis 131 - 19s, 17p, 17d, 14f, 8g}
           \footnotetext{Basis 140 - 20s, 18p, 18d, 15f, 9g}
           \footnotetext{Basis 149 - 21s, 19p, 19d, 16f, 10g}
           \footnotetext{Basis 158 - 22s, 20p, 20d, 17f, 11g}
           \footnotetext{Basis 167 - 23s, 21p, 21d, 18f, 12g}
           \footnotetext{Basis 176 - 24s, 22p, 22d, 19f, 13g}
           \footnotetext{Basis 185 - 25s, 23p, 23d, 20f, 14g}
           \footnotetext{Basis 194 - 26s, 24p, 24d, 21f, 15g}
           \footnotetext{Basis 203 - 27s, 25p, 25d, 22f, 16g}
           \footnotetext{Basis 212 - 28s, 26p, 26d, 23f, 17g}
           \footnotetext{Basis 221 - 29s, 27p, 27d, 24f, 18g}
	        \end{tabular}
            \end{ruledtabular}		           
\label{conv_tab}
\end{sidewaystable*}

\begin{table}
    \caption{Convergence trend of $\alpha$ (a.u.) for ground 
	state, $5s^2\;^1S_{0}$, of Sr from PRCC calculation 
	as function of basis size.}
  \begin{ruledtabular}
  \begin{tabular}{cr}
Basis size  & $\alpha$  \\
            \hline
Basis-68      & $161.1397$  \\
Basis-77      & $170.9020$  \\
Basis-86      & $170.2805$  \\
Basis-95	  & $170.0681$   \\
Basis-104	  & $176.8809$   \\
Basis-113     & $178.7907$  \\
Basis-122     & $175.9380$  \\
Basis-131     & $174.8931$  \\
Basis-140     & $174.1457$  \\
Basis-149     & $174.0953$  \\
        \end{tabular}
        \end{ruledtabular}
\label{conv-alpha}
\end{table}

\bibliography{references}
\end{document}